\documentclass[12]{article}

\usepackage[dvipsnames]{xcolor}
\usepackage{color}
\usepackage{fancyvrb}
\usepackage{amsmath} 
\usepackage{amssymb}
\usepackage{listings}
\usepackage{verbatim}
\usepackage{graphicx}
\usepackage{amsmath}
\usepackage[round]{natbib}

\usepackage{amsmath,amssymb,amsfonts,latexsym,mathrsfs}
\usepackage{amsthm}
\usepackage{fullpage}  
\usepackage{bbm}
\usepackage{fancyvrb}

\author{
Evgeny Levi \\
\and
Radu V. Craiu \thanks{Email: craiu@utstat.toronto.edu} 
}
\title{ Bayesian Inference for Conditional Copulas  using  Gaussian Process Single Index Models}
\date{\small Department of Statistical Sciences, University of Toronto}

\usepackage{geometry}
\newcommand{\beq}{\begin{equation}}
\newcommand{\eeq}{\end{equation}}
\newcommand{\beqn}{\begin{eqnarray}}
\newcommand{\eeqn}{\end{eqnarray}}

\newcommand{\RR}{{\mathbf R}}

\newcommand{\rcov}{\mbox{Cov}}
\newcommand{\rcorr}{\mbox{Corr}}
\newcommand{\ru}{{\mbox{Uniform}}}

\begin{document}

\maketitle

\begin{abstract}
Parametric conditional copula models allow the copula parameters to vary with a set of  covariates according to an unknown calibration function.    Flexible Bayesian inference for the calibration function of a
bivariate conditional copula is proposed via a sparse Gaussian process  (GP) prior distribution over the set of smooth calibration  functions for the single index model (SIM). The estimation of parameters from the marginal distributions and the 
calibration function is done jointly via  Markov Chain Monte Carlo sampling from the full posterior distribution. A  new Conditional Cross Validated Pseudo-Marginal  (CCVML) criterion  is introduced in order  to perform copula selection  and is modified using a permutation-based procedure to  assess data support for the simplifying assumption. The  performance of the estimation method and model selection  criteria is studied via a series of simulations using correct and misspecified models with Clayton, Frank and Gaussian copulas and  a numerical application involving  red wine features.
\end{abstract}


\section{Introduction and Motivation}

Copulas are useful in modelling the dependent structure in the data when there is interest in separating it from the marginal models or when none of the  existent multivariate distributions  are  suitable.  For continuous multivariate distributions, the elegant result of  \cite{Sklar:1959} guarantees the existence and uniqueness of the copula $C:[0,1]^{p}\rightarrow [0,1]$ that links the marginal cumulative distribution functions (cdf) and the  joint cdf. Specifically,   $$H(Y_1,\ldots,Y_p)=C(F_1(Y_1),\ldots,F_p(Y_p)),$$ where $H$ is the joint cdf, and $F_{i}$ is the marginal cdf for variable $Y_{i}$,  for $1\le i \le p$, respectively. The extension to conditional distributions via the {\it conditional copula} was  used by \cite{lamb-van} and subsequently formalized by \cite{patton2006modelling}  so that 
\begin {equation}
H(Y_1,\ldots,Y_p|X)=C_X(F_{1|X}(Y_1|X),\ldots,F_{p|X}(Y_p|X)),
\label{def-cc}
\end{equation}
where $X\in R^{q}$ is a vector of conditioning variables, $C_X$ is the conditional copula that may change with $X$ and $F_{i|X}$ is the conditional cdf of $Y_i$ given X for $1\le i\le p$.    A parametric model for the conditional copula assumes $C_X=C_{\theta(X)}$ belongs to a family of copulas and only the  parameter $\theta \in \Theta$
 varies as a function of $X$. In the remaining of this paper we assume that there exists a known 
one-to-one function $g:\Theta \rightarrow \RR$ such that $\theta(X)=g^{-1}(\eta(X))$ with the calibration function $\eta : \RR \rightarrow \RR$ in the inferential focus.

There are a number of reasons one is interested in estimating  the conditional copula.  First,  in regression models with multivariate  responses, one may want to  determine how the dependence structure among the components of the response varies with the covariates. This model will ultimately impact the performance of  model-based prediction.  For instance, for a bivariate response in which one component is predicted given the other, the conditional density takes the form
\beq
 h(y_{1}|y_{2},x) =  f(y_{1}|x)  c_{\theta(x)}( F_{1|x}(y_{1}|x),  F_{2|x}(y_{2}|x)),
 \label{reg-pred}
 \eeq
where $c_{\theta(x)}$ is the density  of the conditional copula $C_{\theta(x)}$. Hence, in addition to the information contained in the  marginal model,  in equation \eqref{reg-pred} we use  for prediction also the information in the other responses. 

Second, when specifying  a general multivariate distribution, the conditional copula is an essential ingredient. For instance, if $U_{1},U_{2},U_{3}$ are three  $\ru(0,1)$ variables then their joint density is
$$c(u_{1},u_{2},u_{3})=c_{12}(u_{1},u_{2}) c_{23}(u_{2},u_{3}) c_{\theta(u_2)}(P(U_1\le u_{1} | u_{2}), P(U_3\le u_{3}|u_2)).$$

Finally, a conditional copula with predictor values $X\in \RR^q$ in which $\eta(X)$ is constant, may exhibit non-constant patterns when some of the components of $X$ are not included in the model. This point will be revisited in section \ref{cop-mis}.

When estimation for the conditional copula model is contemplated, one must consider that there are multiple  sources of error and each will have an impact on the model. Even in the simple case in which the estimation of the marginals and copula suffer from errors that depend only on $x$  one obtains
\begin{align}
&c_{\theta(x)+\delta_{3}(x)}(F_{1|x}(y_{1}|x)+\delta_{1}(x), F_{2|x}(y_{2}|x)+\delta_{2}(x))=c_{\theta(x)}(F_{1|x}(y_{1}|x), F_{2|x}(y_{2}|x)) \label{correct} \\
&+ c^{(1,0,0)}_{\theta(x)}(F_{1|x}(y_{1}|x), F_{2|x}(y_{2}|x))  \delta_{1}(x)  \label{error1}\\
& +  c^{(0,1,0)}_{\theta(x)}(F_{1|x}(y_{1}|x), F_{2|x}(y_{2}|x))  \delta_{2}(x) \label{error2}\\
&+c^{(0,0,1)}_{\theta(x)} (F_{x}(y_{1}),G_{x}(y_{2}))  \delta_{3}(x) + {\cal{O}}(||\delta(x)||^{2}), \label{error3}
\end{align}
where $c^{(1,0,0)},c^{(0,1,0)}$ and $c^{(0,0,1)}$ are the partial derivatives of $c_z(x,y)$ w.r.t. $x$, $y$ and $z$, respectively.
The right hand term in equation \eqref{correct} marks  the correct contribution to the joint likelihood while   \eqref{error1}-\eqref{error3} show the biases incurred due to errors in estimating the first and second marginal conditional cdf's and the copula calibration function, respectively. It becomes apparent that in order to keep the estimation error low, one must consider flexible models for the marginals and the copula.

%

Depending on the strength of assumptions we are willing to make about $\eta(X)$, a number of possible approaches are available. The most direct is to assume a known parametric form for the calibration function, e.g. constant or linear,
and estimate the corresponding parameters by maximum likelihood estimation \citep{genest1995semiparametric}.  This approach relies on knowledge about the shape of the calibration function which, in practice, can be unrealistic. A more flexible approach uses non-parametric methods
\citep{acar2011dependence,vog} and estimate the calibration function using  smoothing methods. For univariate  $X$,  \cite{craiu2012mixed} devised  Bayesian inference based on a flexible cubic spline model and for multivariate $X$, \cite{sabeti2014additive}, \cite{gam-cc} and \cite{klein} avoid the curse of dimensionality that appears even for moderate values of $q$, say $q>3$,  by specifying an additive model structure for the calibration function.  Few alternatives to the  additive structure exist.   One exception is   \cite{hernandez2013gaussian} who used a sparse Gaussian Process (GP) prior for estimating the calibration function  and subsequently used the same construction for vine copulas estimation in \cite{lopez2013gaussian}. 
However, when the dimension of the predictor space is even moderately large the curse of dimensionality prevails and it is  expected that the  $q$-dimensional GP  used for calibration estimation  will not capture important patterns for sample sizes that are not very large.  Moreover,  the full efficiency of the method proposed in \cite{hernandez2013gaussian} is difficult to assess since their model is build with uniform marginals, which in a general setup is equivalent to assuming exact knowledge about the marginal distributions. In fact, when the marginal distributions are estimated it is of paramount importance to  account for the resulting  variance inflation due to error propagation in the copula estimation as reflected by equations \eqref{correct}-\eqref{error3}. The Bayesian model in which joint and marginal components are simultaneously considered will appropriately handle error propagation as long as it is possible to study the full posterior distribution of all the parameters in the model, be they involved in the marginals or copula specification.

Great dimension reduction of the parameter space is achieved  under the so-called {\it simplifying assumption (SA)} that assumes 
$C_X= C$ for all $X$, i.e. the conditional copula is constant \citep{gijbels2015estimation}. The SA condition can  significantly simplify the vine copula estimation \citep[for example, see][]{czado}, but it is known to lead to bias when the model ignores its violation 
\citep{agn}.  Therefore, for conditional copula models it is of practical interest to assess whether the data supports or not SA. A first step towards a formal test for SA can be found in \cite{acar2013statistical}. The reader is referred to \cite{derumigny2016tests} for an excellent review of work on SA, and  ideas for future work.

This paper's contribution  is two-fold: on one hand we consider Bayesian joint analysis of the marginal and copula models using flexible GP models. Our emphasis is placed on the estimation of the calibration function $\eta(x)$ which is assumed to have a GP prior that is  evaluated at $\beta^T X$ for some normalized $\beta$, thus coupling the GP-prior construct  with the {\it single index model  (SIM)} of 
\cite{choi2011gaussian} and \cite{gramacy2012gaussian}.  The GP-SIM is more flexible
than a canonical linear model and computationally more manageable than a full GP with $q$ variables. The proposed model  can be used for large  covariate dimension $q$ and for large samples. Both marginal means will be fitted using  sparse GP approaches so that large data sets can be computationally manageable. The dimension reduction of the SIM approach has been noted also by \cite{ferm2015}, but their method  differs in fundamental aspects  from the one proposed here. 
So far, GP-SIM's have been used mostly in regression settings
where the algorithm of \cite{gramacy2012gaussian}  can be used to efficiently sample the posterior distribution. However,  the GP-SIM model for conditional copulas involves   a non-Gaussian likelihood which requires a new sampling algorithm. 

A second contribution of the paper deals with model selection issues that are particularly relevant for the conditional copula construction.  We consider of importance the choice of copula family and identifying whether the simplifying assumption (SA) is supported by the data.  For the former task we develop a conditional cross-validated marginal likelihood (CCVML) criterion and also examine the performance of the Watanabe Information Criterion \citep{watanabe2010asymptotic}, while for determining whether the data supports the SA assumption or not we construct a 
permutation-based variant of the CVML that shows good performance in our numerical experiments.
Finally, we identify an important  link between SA and missing covariates in the conditional copula model. To our knowledge, this connection has not been reported elsewhere.

In the next section we review the GP-SIM formulation and introduce the notation.  The construction of the  conditional copula model, the computational algorithm and the model selection procedures are covered in Section 3. In Section 4 we illustrate the efficiency of the method via simulation and a numerical analysis of wine data. All the contributions  relevant to the important issue of SA are included in Section 5. The paper ends with conclusions and directions for future work.

\section{Brief review of Bayesian inference for Sparse GP}
Assume we observe  $n$
independent observations $\{(x_i,y_i), \hspace{1ex} i=1\ldots n\}$,
where $Y_i \in \mathbf R$ is the response variable and $X_i \in \mathbf R^q$ is a vector of covariates. Suppose that the probability distribution of $Y_i$ has a known form and depends on $X_i$ through some unknown function $f$, e.g. $Y_i  \stackrel{indep}{\sim} N(f(X_i),\sigma^2)$.
The goal is to estimate the unknown smooth function $f:\mathbf R^q
\rightarrow \mathbf R$.
A Gaussian Process (GP) prior on  the function $f$  implies
\begin{equation}
 (f(X_1),f(X_2),\ldots,f(X_n))^T \sim \mathcal N(0,K(X,X;\mathbf w)),
\end{equation}
where $\mathcal N$$(\mu,\Sigma)$ denotes a multivariate normal distribution with mean $\mu$ and variance covariance matrix $\Sigma$ and $K$ is a covariance matrix which depends on $X_1,\ldots,X_n$ and additional parameters. In this paper we use the squared exponential kernel to model the matrix $K(X,X;\mathbf w)$, i.e. its  $(i,j)$ element is
\begin{equation}
k(X_i,X_j;\mathbf{w})=e^{w_0} \exp\left[-\sum_{s=1}^{s=q}\frac{(X_{is}-X_{js})^2}{e^{w_s}}\right].
\label{gp-cov}
\end{equation}
The unknown parameters $\mathbf{w}=(w_0,\ldots,w_q)$ that determine the strength of dependence in \eqref{gp-cov} are inferred from the data. 


In the case in which the covariate dimension, $q$, is moderately large,  an accurate estimation of $f$ will require a large sample size, $n$. Unfortunately, this desideratum is hindered by the computational complexity involved in fitting a GP model when  $n$ is large, as the MCMC sampler designed to sample from the posterior require at each iteration
 the  calculation and inversion of the matrix $K(X,X;\mathbf w) \in \mathbf R^{n\times n}$. 
   
To make GP models applicable for larger data we follow the literature on {\it sparse GP} \citep[more details can be found in][]{quinonero2005unifying,snelson2005sparse,naish2007generalized} in which it is assumed that  learning about $f$ can be achieved using a smaller sample of $m$ latent variables, called {\it inducing variables}, that channel the information contained in the covariates $\{x_1,\ldots, x_n\}$. To complete the description we consider the following notation: if  the original covariates are 
$X= (x_1,\ldots,x_n)^T \in \mathbf{R}^{n \times q}$  and the inducing variables are 
$X^*= (x_1^*,\ldots,x_{m}^*)^T \in \mathbf{R}^{m\times q}$ 
then we  denote   $K(X,X^*;\mathbf w) \in \mathbf{R} ^{n \times m}$ the matrix
\begin{equation}
K(X,X^*;\mathbf w)=
\begin{bmatrix}
k(x_1,x^*_1;\mathbf w) & \cdots & k(x_1,x^*_{m};\mathbf w) \\
\vdots & \ddots & \vdots \\
k(x_n,x^*_1;\mathbf w) & \cdots & k(x_n,x^*_{m};\mathbf w)
\end{bmatrix},
\end{equation}
where $k(x_{i},x_{j}^{*};\mathbf w)$ is defined as in \eqref{gp-cov}.
We also define the following two matrices that  will be used
throughout the paper
\begin{align}
A(X^*,X;\mathbf w)= & K(X^*,X,\mathbf w)K(X,X,\mathbf w)^{-1} \label{eq:A}, \\
B(X^*,X;\mathbf w)= & K(X^*,X^*,\mathbf w)-K(X^*,X,\mathbf w)K(X,X,\mathbf w)^{-1}K(X^*,X,\mathbf w)^T. \label{eq:B}
\end{align}
The ratio $m/n$ influences the  trade-off between computational efficiency and statistical efficiency, as a smaller $m$ will favour the former and a larger $m$ will ensure no significant loss of the latter. We assume  $X_j^*=X_{u_j}$ for $j=1,\ldots, m$ and the function values for the inducing points are $\mathbf {\tilde f}=(f(x_{u_1}),\ldots,f(x_{u_m}))^T=
(f_{u_1},\ldots,f_{u_m})^T$. 
The joint density of  the response $Y$, the latent variable $\mathbf {\tilde f}$ and the parameter $\mathbf w$ can be expressed
only in terms of the $m$-dimensional vector $\mathbf {\tilde f}$ since
\begin{equation}
P(Y,\mathbf{\tilde f},\mathbf{w}|X,\tilde X)=  P(Y|A(X,\tilde X;\mathbf w)\mathbf {\tilde f}) \mathcal N(\mathbf {\tilde f};0,K(\tilde X,\tilde X;\mathbf{w}))p(\mathbf w),
\end{equation}
where $p(\mathbf{w})$ is the prior probability for the parameters $\mathbf{w}$.
The posterior distribution $\pi(\mathbf {\tilde f},\mathbf{w}|\mathcal D)$
is still  not tractable, but sampling from it will  be much less expensive since  $K(X,\tilde X;\mathbf w) \in \RR^{n\times m}$  and $K(\tilde X,\tilde X;\mathbf w) \in \RR^{m\times m}$.
So far we have assumed that the inducing inputs $\tilde X$ are selected from  the samples collected. 
A data-driven alternative is to choose  
$x_{u_1},\ldots,x_{u_m}$ as the centers of $m$ clusters created from the  original
covariates $X$ via  a simple k-means algorithm \citep{bishop2006pattern}. Intuitively, it makes sense to have more inducing points in  regions that exhibit more variation in covariate values.

Finally, in order to reduce the dimensionality of the parameter space,  we assume 
that
\begin{equation}
\tilde{ f}(x_i)=\tilde{ f}(x_i^T \beta) ,
\label{sim}
\end{equation}
and we set $ \tilde{\mathbf f}=(\tilde{ f}(x_1^T \beta),\ldots, \tilde{ f}(x_n^T \beta))^T$, 
where $\tilde f:\mathbf R\rightarrow \mathbf R$ is an unknown function that is part of inferential focus and $\beta \in \mathbf R^q$ is  normalized, i.e. $\|\beta \| = 1$. Note that without normalization the parameter 
$\beta$ is not identifiable.  
The {\it single index model (SIM)} defined by \eqref{sim} coupled with the sparse GP approach has  the advantage that it casts the original  problem of estimating a general function ${ f}$ in $q$ dimensions  based on $n$ observations into the   estimation  of  $q$-dimensional  parameter vector $\beta$ and   
of the one-dimensional map $\tilde { f}$ based on $m<<n$ inducing points. 
%
%

\section{ GP-SIM for Conditional copula}
Suppose that the observed data $\mathcal D=\{(Y_{1i},Y_{2i},x_i)\hspace{1ex} i=1 \ldots n\}$ 
consists of triplets  $(Y_{1i},Y_{2i},X_i)$ where 
$Y_{1i},Y_{2i}\in \mathbf R$ and $X_i \in \mathbf R^q$.
For notational convenience let $Y_1=(Y_{11},\ldots,Y_{1n})^T$,
$Y_2=(Y_{21},\ldots,Y_{2n})^T$ and $X=(X_1,\ldots,X_n)^T$.
We assume that the marginal distribution of $Y_{ji}$ ($j=1,2$)
is Gaussian with mean $f_{j}(x_i)$ and constant variance
$\sigma^2_{j}$. If we let $\mathbf f_j=(f_j(x_1),\ldots,f_j(x_n))^T$
we can compactly write:
\begin{equation}
P(Y_j|X)=\mathcal N(\mathbf f_j,\sigma^2_j \mathbf I_n)\hspace{2ex} j=1,2.
\end{equation}
We use a conditional copula to account for the fact that the dependence between the responses varies with covariate $X$.  The likelihood is
\begin{equation}
\begin{split}
P(Y_1,Y_2|X)=&\prod_{i=1}^{n}\frac{1}{\sigma_1}\phi\left(\frac{Y_{1i}-\mathbf f_{1i}}{\sigma_1}\right)
\frac{1}{\sigma_2}\phi\left(\frac{Y_{2i}-\mathbf f_{2i}}{\sigma_2}\right)\times\\
&\times c\left(\Phi\left(\frac{Y_{1i}-\mathbf f_{1i}}{\sigma_1}\right),\Phi\left(\frac{Y_{2i}-\mathbf f_{2i}}{\sigma_2}\right)|\theta(x_i)\right).
\end{split}
\end{equation}
Here $c$ denotes a parametric copula density function, while $\Phi$ and $\phi$ are the cumulative
probability function and density function of a standard normal distribution, respectively.  The parameter of a copula depends
on the unknown function $\theta(x_i)=g^{-1}(f(x_i))$, where $f$ is assumed to take the form given in \eqref{sim} and $g$ is a known invertible link
function that allows an unrestricted parameter space for $\mathbf{f}$.  It is worth noting that the GP-SIM model used for estimating the copula parameter is invariant to the scale used. For instance, whether one chooses to estimate the calibration for the copula parameter, $\theta(X)$, or   Kendall's  $\tau(X)$, the form given in \eqref{sim} will be valid for both. However, this is not true in general for additive models, since a non-linear transformation will break the additivity.  

The GP-SIM is fully specified once we assign the GP priors to $f_1,f_2,f$ and the parametric priors for the remaining parameters, as follows:
\begin{equation}
\begin{split}
f_1 \sim & \hspace{1ex} \mathcal {GP}(\mathbf w_1), \;\;
f_2 \sim  \hspace{1ex} \mathcal {GP}(\mathbf w_2), \; \;
f \sim   \hspace{1ex}\mathcal {GP}(\mathbf w),    \\
\mathbf w_1 \sim & \hspace{1ex} \mathcal N(0,5\mathbf I_{q+1}),\;\;
\mathbf w_2 \sim  \hspace{1ex} \mathcal N(0,5\mathbf I_{q+1}),\;\;
\mathbf w \sim  \hspace{1ex} \mathcal N(0,5\mathbf I_{2}), \\
\beta \sim & \hspace{1ex} \mbox{U}(S^{q-1}),\;\;
\sigma_1^2 \sim  \hspace{1ex} \mathcal {IG}(0.1,0.1),\;\;
\sigma_2^2 \sim  \hspace{1ex} \mathcal {IG}(0.1,0.1).
\end{split}
\end{equation}
The $\mathcal {GP}(\mathbf w)$ is a Gaussian Process prior
with mean of 0, squared exponential kernel with parameters
$\mathbf w$, U$(S^{q-1})$ is a uniform
distribution on the surface of the $q$-dimensional unit sphere and $\mathcal {IG}(\alpha,\beta)$   denotes the inverse gamma distribution.
Because the focus of the paper is on inference for the copula, we  allow  $f_1$ and $f_2$ to be  evaluated on $\mathbf R^q$ while
$f$ is on $\mathbf R$. In order to avoid computational problems that affect the GP-based inference when the sample size is large, the inference  will rely on the Sparse GP method that was described in the previous section.
Suppose $\tilde X_1$ are $m_1$ inducing inputs for function $f_1$,
$\tilde X_2$ are $m_2$ inducing inputs for function $f_2$ and
$\tilde Z$ are $m$ inducing inputs for function $f$. Also
let $\mathbf{\tilde f}_1$ be $f_1$ evaluated at $\tilde X_1$,
$\mathbf{\tilde f}_2$ be $f_2$ evaluated at $\tilde X_2$ and
$\mathbf{\tilde f}$ be $f$ evaluated at $\tilde Z$.
Then
the joint density of the observed data and parameters is proportional to:
\begin{equation}
\begin{split}
P(Y_1,Y_2,\mathbf{\tilde f}_1,& \mathbf{\tilde f}_2,\mathbf{\tilde f},
\mathbf w_1,\mathbf w_2,\mathbf w,\sigma_1^2,\sigma_2^2,\beta|
X,\tilde X_1,\tilde X_2,\tilde Z)\propto
\mathcal N(Y_1;\mathbf f_1,\sigma^2_1\mathbf I_n)
\mathcal N(Y_2;\mathbf f_2,\sigma^2_2\mathbf I_n)\times\\
\times & \prod_{i=1}^{i=n}
c\left(\Phi\left(\frac{Y_{1i}-\mathbf f_{1i}}{\sigma_1}\right),\Phi\left(\frac{Y_{2i}-\mathbf f_{2i}}{\sigma_2}\right)|g^{-1}(\mathbf f_{i})\right)
\mathcal N(\mathbf {\tilde f}_1;0,K(\tilde X_1,\tilde X_1;\mathbf w_1))\times \\
\times & \mathcal N(\mathbf {\tilde f}_2;0,K(\tilde X_2,\tilde X_2;\mathbf w_2))
\mathcal N(\mathbf {\tilde f};0,K(\tilde Z,\tilde Z;\mathbf w)) 
\mathcal N(\mathbf w_1;0,5\mathbf I_{q+1})\times \\
\times & \mathcal N(\mathbf w_2;0,5\mathbf I_{q+1})
\mathcal N(\mathbf w;0,5\mathbf I_{2})
\mathcal {IG}(\sigma^2_1;0.1,0.1)
\mathcal {IG}(\sigma^2_2;0.1,0.1),
\end{split}
\end{equation}
where
$ \mathbf f_1 = A(X,\tilde X_1;\mathbf w_1) \mathbf{\tilde f}_1$, $\mathbf f_2 = A(X,\tilde X_2;\mathbf w_2) \mathbf{\tilde f}_2$  and $\mathbf f = A(X\beta,\tilde Z;\mathbf w) \mathbf{\tilde f} .
$
The number of inducing inputs $m_1$, $m_2$ and $m$ can all be different 
but in our applications we will choose their values to be significantly smaller than the sample size, $n$. Ideally we need the number of inducing inputs to be as large as possible but at the same time make the MCMC implementation computationally feasible. 

As suggested earlier we can define $\tilde X_1$ and $\tilde X_2$ 
as centers of $m_1$ and $m_2$ clusters of $X$. So if $m_1$ is the same
as $m_2$ then inducing inputs would also be the same. We cannot use
the same strategy for $\tilde Z$, since then we would need
the  centers for the clusters of the variable  $X^T \beta$ which are unknown. 
If we assume that each covariate $x_{ij}$ is between 0 and 1 (this can be achieved easily 
if we subtract the  the minimum value and divide by range) then following the 
Cauchy-Schwartz inequality we obtain
\[\| x_i^T\beta \| \leq \sqrt{\| x_i \|^2
\| \beta \|^2} \leq \sqrt{q} \hspace{2ex} \forall x_i,\beta .\]
Hence we can choose $\tilde Z$ to be $m$ equally spaced points in the interval
$[-\sqrt{q},\sqrt{q}]$. 

The contribution of the conditional copula model to the joint likelihood breaks the tractability of the posterior  conditional densities and complicates the design of an efficient MCMC algorithm that can sample efficiently from the posterior distribution. 
The conditional joint posterior
distribution of the latent variables ($\mathbf f$) and parameters ($\mathbf w$) given the observed data $\mathcal D$ does not have a tractable form and its study will require the use of 
 Markov Chain Monte Carlo (MCMC) sampling methods. 
Specifically,  we use Random Walk Metropolis (RWM) within Gibbs sampling
for $\mathbf w$ \citep{craiu2014bayesian,rosenthal2009markov,andrieu2003introduction} while for $\mathbf f$ we will use the elliptical slice sampling \citep{murray2009elliptical}
that has been designed specifically  for GP-based models and does not require tuning of
free parameters.

\subsection{Computational Algorithm}
Inference is  based on the posterior distribution
$ \pi(\mathbf{\tilde f}_1,\mathbf{\tilde f}_2,\mathbf{\tilde f},
\mathbf w_1,\mathbf w_2,\mathbf w,\sigma_1^2,\sigma_2^2,\beta|
\mathcal D,\tilde X_1,\tilde X_2,\tilde Z)$
which  is not mathematically tractable, so the study of its properties will rely on  Monte Carlo sampling. In this section we provide the detailed steps of the MCMC sampler designed to sample from $\pi$.

The general form of the algorithm falls within the class of  Metropolis-within-Gibbs (MwG) samplers in which we update in turn each component of the chain by sampling from its conditional distribution, given all the other components. The presence of the copula in the likelihood breaks the usual conditional conjugacy of the GP models so none of the
components have  conditional distributions that can be sampled directly. 

Suppose we are interested in sampling a target $\pi(\theta_{1},\ldots,\theta_{k})$. A generic  MwG sampler proceeds as follows:

\begin{itemize}
\item[Step I] Initialize the chain  at $\theta_1^{(1)},\theta_2^{(1)},\ldots,\theta_{k}^{(1)}$.
\item[Step R] At iteration $t+1$ run iteratively the following steps for each $j=1,\ldots,k$:
\begin{enumerate}
\item Sample 
$\theta_{j}^{*}\sim q_{j}(\theta_{j}|\theta_{-j}^{(t+1;t)})$
where $\theta_{-j}^{(t+1;t)}=(\theta_{1}^{(t+1)},\ldots, \theta_{j-1}^{(t+1)},\theta_{j+1}^{(t)},\ldots,\theta_{k}^{(t)})$ is the most recent state of the chain with the first $j-1$ components updated already (hence the supraindex $t+1$), the $j$th component removed and the remaining $n-j$ components having  the values determined at iteration $t$ (hence the supraindex $t$).

\item Compute $r=\min \left \{1, {\pi(\theta_{1}^{(t+1)},\ldots, \theta_{j-1}^{(t+1)}, \theta_{j}^{*},\theta_{j+1}^{(t)},\ldots,\theta_{k}^{(t)}) q_{j}(\theta_{j}^{(t)}|\theta_{-j}^{(t+1;t)}) \over \pi(\theta_{1}^{(t+1)},\ldots, \theta_{j-1}^{(t+1)}, \theta_{j}^{t},\theta_{j+1}^{(t)},\ldots,\theta_{k}^{(t)}) q_{j}(\theta_{j}^{(*)}|\theta_{-j}^{(t+1;t)}) } \right \} $.

\item With probability $r$ accept proposal and set $\theta_{j}^{(t+1)}=\theta_{j}^{*}$ and with $1-r$ reject proposal and let $\theta_{j}^{(t+1)}=\theta_{j}^{(t)}$.

\end{enumerate}

\end{itemize}

The proposal density $q_{j}(\cdot|\cdot)$ corresponds to the transition kernel used for the $j$th  component. Our algorithm uses a number of  proposals corresponding to 
  { Random Walk Metropolis-within-Gibbs} (RWMwG), { Independent Metropolis-within-Gibbs (IMwG)}
 and { Elliptical Slice Sampling
within Gibbs (SSwG)} moves.

At the $t+1$ step we use the following proposals to update the chain: 
\begin{itemize}
\item[$\mathbf w_i$:]  Use a RWM transition kernel:  $\mathbf w^* \sim \mathcal N(\mathbf w_i^{(t)},c_{w_{i}}\mathbf I_{d+1})$.
The constant $c_{w_{i}}$ is chosen so that the acceptance rate is about 30\%, $i=1,2$.
\item[$\mathbf w$:] Use the RWM:  $\mathbf w^* \sim \mathcal N(\mathbf w^{(t)},c_{w}\mathbf I_{2})$. The constant $c_{w}$ is chosen so that the acceptance rate is about 30\%.
\item[$\sigma^2_i$:] Without the copula, the conditional posterior distribution
of $\sigma^2_i$ would be $\mathcal{IG}(0.1+n/2,
0.1+(Y_i-A_i\mathbf {\tilde f}_i^{(t)})^T(Y_i-A_i\mathbf {\tilde f}_i^{(t)}))$
where $A_i=A(X,\tilde X_i;\mathbf w_i^{(t+1)})$ for all $i=1,2$. We will use this 
distribution as a proposal distribution in the IM transition kernel, i.e. the proposal is 
$(\sigma_i^2)^* \sim \mathcal{IG}(0.1+n/2,
0.1+(Y_i-A_i\mathbf {\tilde f}_i^{(t)})^T(Y_i-A_i\mathbf {\tilde f}_i^{(t)}))$.
The acceptance rate is usually in the range of $[0.25,0.60]$ and the chain mixes better than it would under a RWM.
\item[$\beta$:] Since $\beta$ is normalized we will use RWM
on unit sphere using `Von-Mises-Fisher'
distribution (henceforth denoted $\mathcal {VMF}$).
The VMF distribution has two parameters, $\mu$ (normalized to have norm one) which represents the mean direction
and $\kappa$, the concentration parameter. A larger
$\kappa$ implies that the distribution will be more concentrated around $\mu$. The
density is symmetric in $\mu$ and the argument and is proportional to $f_{VMF}(x;\mu,\kappa)\propto
\exp(\kappa x^T \mu)$.

The proposals are generated using  $\beta^* \sim \mathcal {VMF}(\beta^{(t)},\kappa)$, where
 $\kappa$ is chosen so that the acceptance rate is around 30\%.

\item[$\mathbf{\tilde f}$'s:] For $\mathbf{\tilde f}_{i}$, $i=1,2$ and $\mathbf{\tilde f}$ we use the elliptical slice sampling proposed by \cite{murray2009elliptical} which
does not require the tuning of simulation parameters. 
\end{itemize}
\vspace{0.5cm}
In our experience the efficiency of the algorithm benefits from  
initial values that are not too far from the posterior mode. Therefore we propose first to estimate the two
independent
regressions for $Y_1$ and $Y_2$
to get $(\mathbf{\tilde f}_1,\mathbf w_1,\sigma_1^2)^{(1)}$ and
$(\mathbf{\tilde f}_2,\mathbf w_2,\sigma_2^2)^{(1)}$. Then run
another MCMC fixing marginals and only sampling $(\mathbf{\tilde
f},\mathbf w)$. This procedure estimates $(\mathbf{\tilde f},\mathbf w)^{(1)}$.
These 3 short chains (100-200 iterations each) give point-estimates of true parameters and these
estimates can be used as initial values for the joint MCMC. This simple approach shortens the time it would take for the original chain to find the regions of high mass under the posterior. 

Empirically we have also found, that for faster convergence it is better
to start with small $w_1$ values (allowing for more variation in the calibration function). If the chain starts in
large $w_1$ values, it requires a large number of simulations before
it moves to the correct region in the sample space.

\subsection{Model Selection}
The conditional copula model involves two types of selection. First one needs to choose the copula family from  a set of possible candidates.  Second, it is often of interest to determine whether a parametric simple form for the calibration is supported by the data. For instance, a constant calibration function indicates that the dependence structure does not vary with the covariates, a conclusion that may be of scientific interest in some applications.
We investigate the performance of three measures of fit that
can be estimated from the MCMC samples $\mathbf \omega^{(t)}\hspace{1ex}
t=1\ldots M$ 
where $\mathbf \omega^{(t)}$ is the vector of parameters and latent variables drawn at step $t$ from the posterior corresponding to 
 model $\mathcal M$.

\subsection{Cross-Validated Pseudo Marginal Likelihood}
The cross-validated pseudo marginal likelihood (CVML)
\citep{geisser1979predictive, hanson}
calculates the average (over parameter values) prediction power for  model $\mathcal M$ via
\begin{equation}
\mbox{CVML}(\mathcal M)=\sum_{i=1}^{n}\log\left(P(Y_{1i},Y_{2i}|\mathcal D_{-i},\mathcal M)\right),
\label{cvml-def}
\end{equation}
where $\mathcal D_{-i}$ is the data set from which the $i$th observation has been removed.
 An estimate of \eqref{cvml-def} can be obtained using posterior draws for all the parameters and latent variables in the model  \citep[see, for example,][]{sabeti2014additive}. Specifically, if the latter are  denoted by $\mathbf \omega$ , then
\begin{equation}  
E\left[P(Y_{1i},Y_{2i}|\mathbf \omega,\mathcal M)^{-1}\right]=P(Y_{1i},Y_{2i}|\mathcal D_{-i},\mathcal M)^{-1},
\end{equation}
where the expectation is with respect to conditional (posterior) distribution of $\mathbf \omega$
given full data $\mathcal D$ and the model $\mathcal M$.
Based on the posterior samples we can estimate the CVML as
\begin{equation}
\mbox{CVML}_{est}(\mathcal M)=-\sum_{i=1}^{n}\log\left(\frac{1}{M}
\sum_{t=1}^{M}P(Y_{1i},Y_{2i}|\mathbf \omega^{(t)},\mathcal M)^{-1}\right).
\end{equation}
The model with the largest CVML is selected. 

\subsection{Conditional CVML criterion}

The conditional copula construction is particularly useful in  predicting one response given the other ones. We exploit this feature by computing the predictive distribution of one response given the rest of the data. The resulting {\it conditional CVML (CCVML)}  is computed from the  $P(Y_{1i}|Y_{2i},\mathcal D_{-i})$ and $P(Y_{2i}|Y_{1i},\mathcal D_{-i})$ via
\begin{equation}
\mbox{CCVML}(\mathcal M)=\frac{1}{2}\left\{ \sum_{i=1}^{n}\log\left[P(Y_{1i}|Y_{2i},\mathcal D_{-i},\mathcal M)\right] + \sum_{i=1}^{n}\log\left[P(Y_{2i}|Y_{1i},\mathcal D_{-i},\mathcal M)\right] \right\}.
\end{equation}
Note that when the marginal distributions are uniform, then
CCVML is the same as CVML. Using a technique similar to the one used in \cite{sabeti2014additive} one can show that
\begin{equation}
\begin{split}  
& E\left[P(Y_{1i}|Y_{2i},\mathbf \omega,\mathcal M)^{-1}\right]=E\left[\frac{P(Y_{2i}|\mathbf \omega,\mathcal M)}{P(Y_{1i},Y_{2i}|\mathbf \omega,\mathcal M)}\right]=P(Y_{1i}|Y_{2i},\mathcal D_{-i},\mathcal M)^{-1} ,\\
& E\left[P(Y_{2i}|Y_{1i},\mathbf \omega,\mathcal M)^{-1}\right]=E\left[\frac{P(Y_{1i}|\mathbf \omega,\mathcal M)}{P(Y_{1i},Y_{2i}|\mathbf \omega,\mathcal M)}\right]=P(Y_{2i}|Y_{1i},\mathcal D_{-i},\mathcal M)^{-1}.
\end{split}
\label{trick}
\end{equation}
Based on \eqref{trick} one can  easily estimate $\mbox{CCVML}$ from MCMC samples:
\begin{equation}
\mbox{CCVML}_{est}(\mathcal M)=-\frac{1}{2}\sum_{i=1}^{n}\left\{\log\left[\frac{1}{M}
\sum_{t=1}^{M}\frac{P(Y_{2i}|\mathbf \omega^{(t)},\mathcal M)}{P(Y_{1i},Y_{2i}|\mathbf \omega^{(t)},\mathcal M)}\right]
+\log\left[\frac{1}{M}
\sum_{t=1}^{M}\frac{P(Y_{1i}|\mathbf \omega^{(t)},\mathcal M)}{P(Y_{1i},Y_{2i}|\mathbf \omega^{(t)},\mathcal M)}\right] \right\}.
\end{equation}

%

\subsection{Watanabe-Akaike Information Criterion}
The Watanabe-Akaike Information Criterion \citep[WAIC,][]{watanabe2010asymptotic} is an information-based criterion that is closely related to the CVML \citep[see][for a dicussion of the connection between CVML and WAIC]{gelman2014understanding}.   

The WAIC is defined as 
\begin{equation}
\mbox{WAIC}(\mathcal M)=-2\mbox{fit}(\mathcal M) + 2\mbox{p}(\mathcal M)  ,
\label{waic}
\end{equation}
where the model fitness is 
\begin{equation}
\mbox{fit}(\mathcal M)=\sum_{i=1}^{n}\log E\left[P(y_{1i},y_{2i}|\mathbf \omega,\mathcal M)\right]
\label{eq:fitness}
\end{equation}
and the penalty 
\begin{equation}
\mbox{p}(\mathcal M)=\sum_{i=1}^{n}Var[ \log P(y_{1i},y_{2i}|\mathbf \omega,\mathcal M)]. 
\label{eq:penalty}
\end{equation}
The expectation in \eqref{eq:fitness} and the variance in \eqref{eq:penalty} are with respect to the conditional  distribution of  $\omega$ given the data and can be computed using the samples produced by the MCMC sampler that draws from $\pi$.
For instance, the Monte Carlo estimate of the fit is 
\begin{equation}
\widehat{\mbox{fit}}(\mathcal M)=\sum_{i=1}^{n}\log\left( \frac{\sum_{t=1}^{M}P(y_{1i},y_{2i}|\mathbf \omega^{(t)},\mathcal M)}{M}   \right),
\end{equation}
and  $\mbox{p}(\mathcal M)$ can be estimated similarly using  the posterior samples 
The model with the smallest WAIC is preferred. In the next section we also investigate via simulations the performance of  CVML, CCVML and WAIC criteria  when identifying data support for a constant calibration function. 
\section{Performance of the algorithms}
\subsection{Simulations}
The  purpose of the simulation study is to assess empirically: 1) the performance of the estimation method under the correct and misspecified models, as well as  2) the ability of the model selection criteria to identify the correct copula structure, i.e. the copula family and the parametric form of the calibration function. For the former aim we compute the integrated mean square for various quantities of interest, including the Kendall's $\tau$.
In order to facilitate estimation performance across different copula families, we  estimate the calibration function on the Kendall's $\tau$  scale. The latter is given by
$$
\tau(X)=4\left(\iint C(U_1,U_2|X)c(U_1,U_2|X)dU_1dU_2\right)-1.
$$
We will compare 3 copulas: Clayton, Frank and  Gaussian under the general GP-SIM model and the Clayton with constant calibration function.
To fit the model with constant copula, we still use MCMC but instead of $\mathbf f, \mathbf{\tilde f}, \mathbf w$
and $\beta$ in calibration we have a constant scalar copula parameter, $\theta$. The RWMwG transition is used
to sample $\theta$, as the proposal distributions for marginals' parameters and latent variables remain the same.

Table~\ref{table:links} provides inverse-link functions $g^{-1}$ used for calibration and the functional relationship between
Kendall's $\tau$ and copula parameters.
\begin{table} [!ht]
\begin{center}
\begin{tabular}{l| l l }
Copula      & Inv-Link function                  & Kendall's $\tau$ formula \\  \hline
Clayton     & $\theta=\exp(f)-1$                 & $\tau=\frac{\theta}{\theta+2} $\\
Frank       & $\theta=f$                         &  No closed form      \\
Gaussian, T & $\theta=\frac{\exp(f)-1}{\exp(f)+1}$ &  $\tau=\frac{2}{\pi}\arcsin{\theta}  $   \\
Gumbel      & $\theta=\exp(f)+1$                 & $1-\frac{1}{\theta}$                   \\
\hline
\end{tabular}
\caption{Inverse-link functions and the functional relationship between Kendall's $\tau$ and the copula parameter. }
\label{table:links}
\end{center}
\end{table}
In addition of Kendall's $\tau$ we use also the conditional mean of $Y_1$ given  $Y_2$ and $X$ for assessing the estimation.   Such conditional means can be useful in prediction when one of the responses is more expensive to measure than the other. The calculation is mathematically straightforward 
\begin{equation}
E(Y_1|Y_2=y_2,X=x)=f_1(x)+\sigma_1\int_0^1\Phi^{-1}(z)c\left(z,
\Phi\left(\frac{y_2-f_2(x)}{\sigma_2}\right);\theta(x)\right)dz.
\label{cexp1}
\end{equation}
If we assume that marginal distributions are uniform then we have the simpler
expression:
\begin{equation}
E(U_1|U_2=u_2,X=x)=\int_0^1c(z,u_2;\theta(x))dz.
\label{cexp2}
\end{equation}
The integrals  in \eqref{cexp1} and \eqref{cexp2} are usually not tractable, but can be easily estimated
via numerical integration since they are one-dimensional and defined on the
closed interval $[0,1]$.

\subsection{Simulation Details}
We generate samples of size  $n=400$  from each of
the next 6 scenarios using the Clayton copula. The covariates are 
generated independently from $\ru(0,1)$ distribution. The covariate dimension $q$
in Scenario 3 is 10, in all other scenarios it is 2. 
\begin{itemize}
\item[\textbf{Sc1}] $f_1(x)=0.6\sin(5x_1)-0.9\sin(2x_2)$,\\
           $f_2(x)=0.6\sin(3x_1+5x_2)$,\\
           $\tau(x)=0.7+0.15\sin(15x^T\beta)$\\
           $\beta=(1,3)^T/\sqrt{10}$,   $\sigma_1=\sigma_2=0.2$ 
                  
\item[\textbf{Sc2}] $f_1(x)=0.6\sin(5x_1)-0.9\sin(2x_2)$\\
           $f_2(x)=0.6\sin(3x_1+5x_2)$\\
            $\tau(x)=0.3\sin(5x^T\beta)$ \\
           $\beta=(1,3)^T/\sqrt{10}$, $\sigma_1=\sigma_2=0.2$

\item[\textbf{Sc3}] $\beta=(1,10,-3,6,1,-6,3,7,-1,-5)^T/\sqrt{267}$, $\sigma_1=\sigma_2=0.2$\\
           $f_1(x)=\cos(x^T\beta)$\\
           $f_2(x)=\sin(x^T\beta)$\\
           $\tau(x)=0.7+0.20\sin(5x^T\beta)$ 
           
\item[\textbf{Sc4}] $f_1(x)=0.6\sin(5x_1)-0.9\sin(2x_2)$\\
           $f_2(x)=0.6\sin(3x_1+5x_2)$\\
           $\tau(x)=0.5$\\
           $\sigma_1=\sigma_2=0.2$ 
           
\item[\textbf{Sc5}] $f_1(x)=0.6\sin(5x_1)-0.9\sin(2x_2)$\\
           $f_2(x)=0.6\sin(3x_1+5x_2)$\\
           $\eta(x)=1+0.7\sin(3x_1^3)-0.5\cos(6x_2^2)$\\
           $\sigma_1=\sigma_2=0.2$ 

\item[\textbf{Sc6}] $f_1(x)=0.6\sin(5x_1)-0.9\sin(2x_2)$\\
           $f_2(x)=0.6\sin(3x_1+5x_2)$\\
           $\eta(x)=1+0.7x_1-0.5x_2^2$\\
           $\sigma_1=\sigma_2=0.2$ 
\end{itemize}
   \textbf{Sc1} and \textbf{Sc2} have calibration functions for which the SIM model is true  for Kendall's
$\tau$ and, consequently, also  for the copula parameter.  \textbf{Sc1} 
corresponds to large dependence ($\tau$ greater than $0.5$) while
\textbf{Sc2} has small dependence ($\tau$ is between $-0.3$ and $0.3$). \textbf{Sc3} also has SIM form for 
calibration function the covariate dimension is $q=10$, so this scenario is important to evaluate how well the algorithms
scale up with dimension. 
 \textbf{Sc4}
corresponds to the covariate-free dependence ($\tau=0.5$) and allows us to verify the  power to detect simple parametric forms for the calibration. Scenarios \textbf{Sc5} and \textbf{Sc6} do not have SIM form,
but have additive calibration function  \citep[as in][]{sabeti2014additive}.
They will be useful to evaluate the effect of 
model misspecification on the inference. Note that \textbf{Sc6} has almost SIM calibration
when $x_2\in [0,1]$. 
For all scenarios we use $m=30$ inducing inputs for all the sparse GP procedures (marginals and copula). \\
The MCMC samplers were run for 40000 iterations for \textbf{Sc3}, and 10000 iterations for all other
scenarios. Simulations for {\bf Sc3} require larger Monte Carlo runs because the parameter space is 32-dimensional compared to 8-dimensional in all other scenarios. The first half of the MCMC sample is discarded as burn-in and the second half is used for inference. As noted earlier,  starting values were found by running two GP regressions separately to estimate marginal
parameters and one MCMC sampler was run in order to estimate calibration parameters. All three samplers were run for only 100 iterations.

\subsubsection{Proof of concept based on one Replicate}

\begin{figure}[!ht]
\begin{center}
\includegraphics[width=16cm]{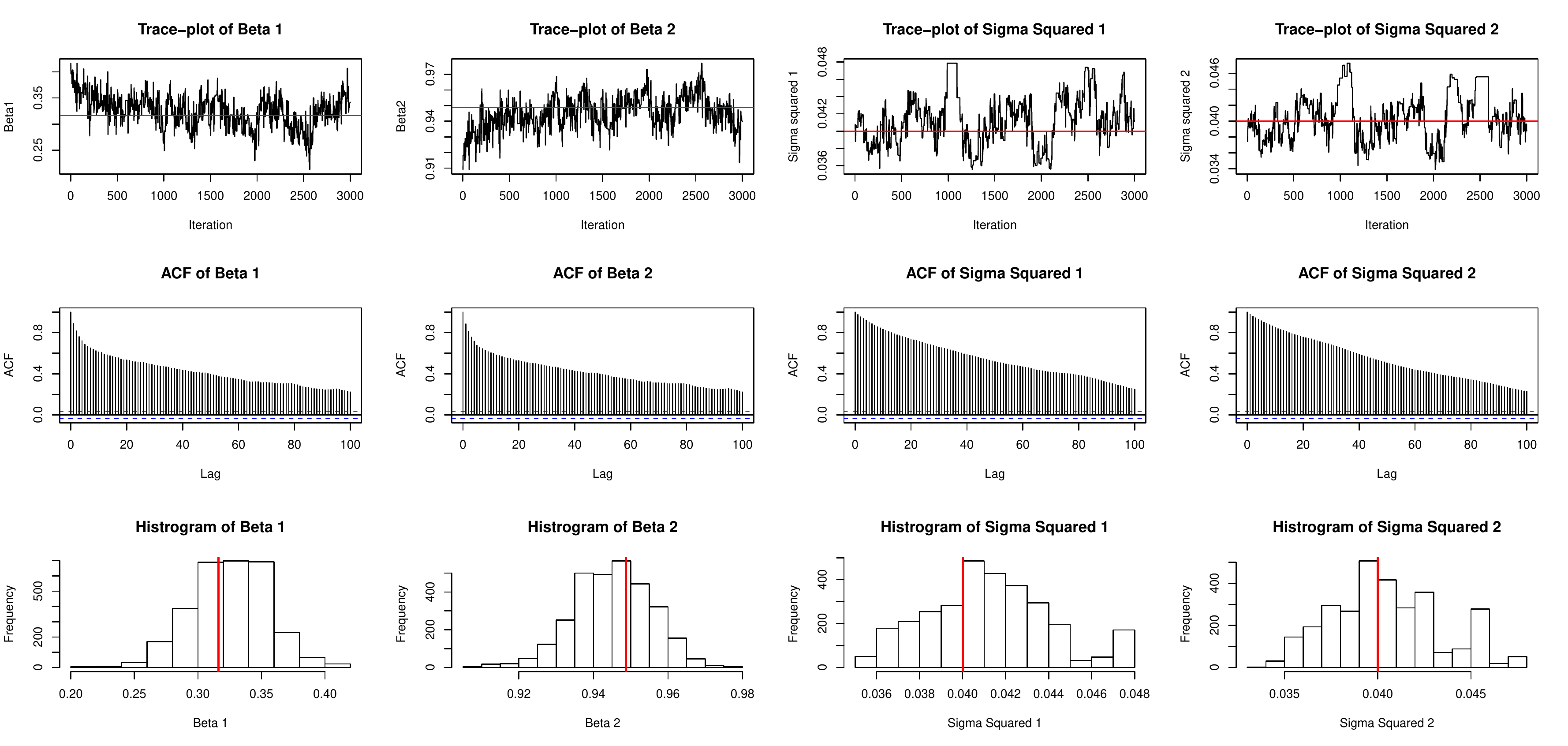}
\caption {
\textbf{Sc1}: Trace-plots, ACFs and histograms of parameters based on  MCMC  samples generated under the true Clayton family. }
\label{fig:part1_trace}
\end{center}
\end{figure}

\begin{figure}[!ht]
\begin{center}
\includegraphics[height=5cm,width=7cm]{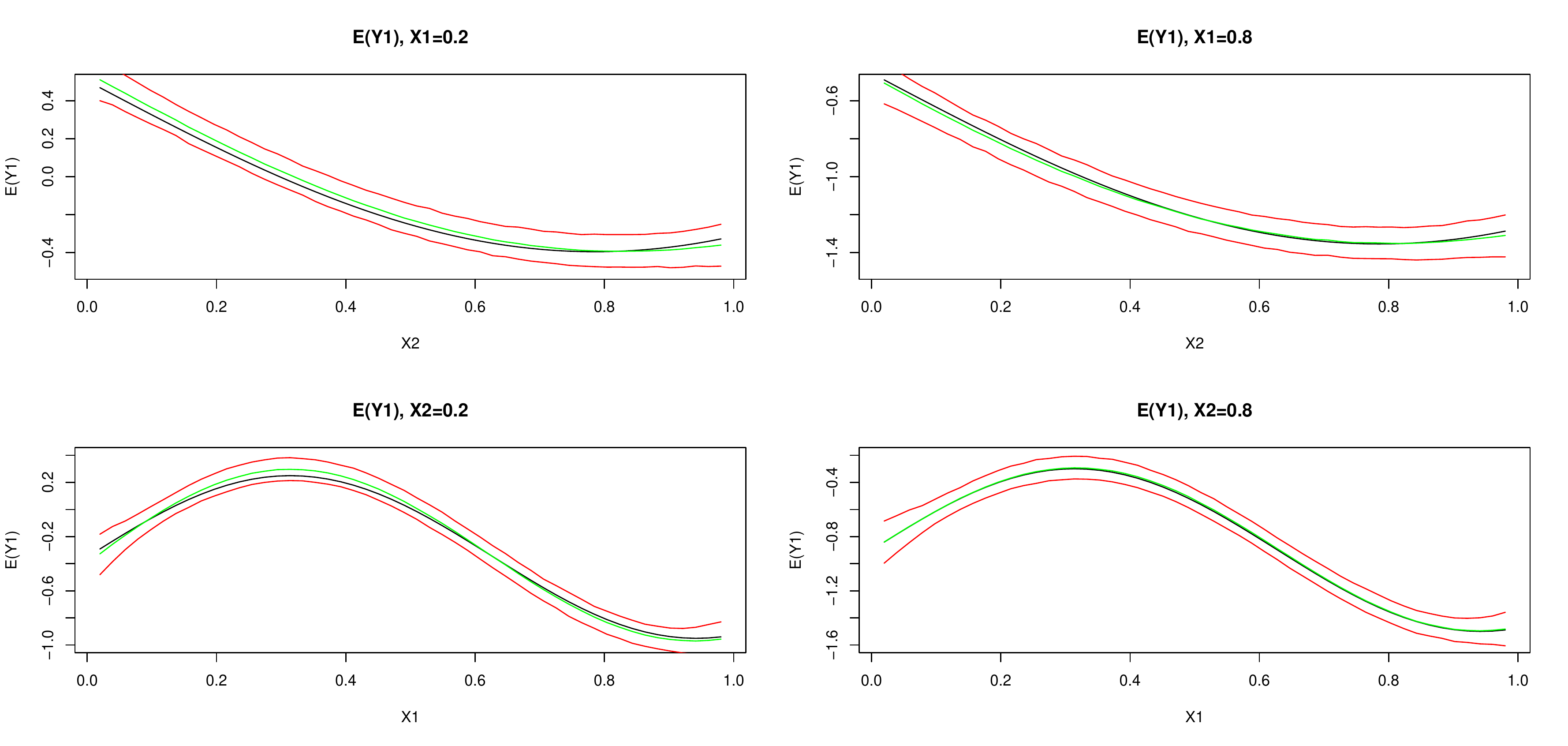}
\includegraphics[height=5cm,width=7cm]{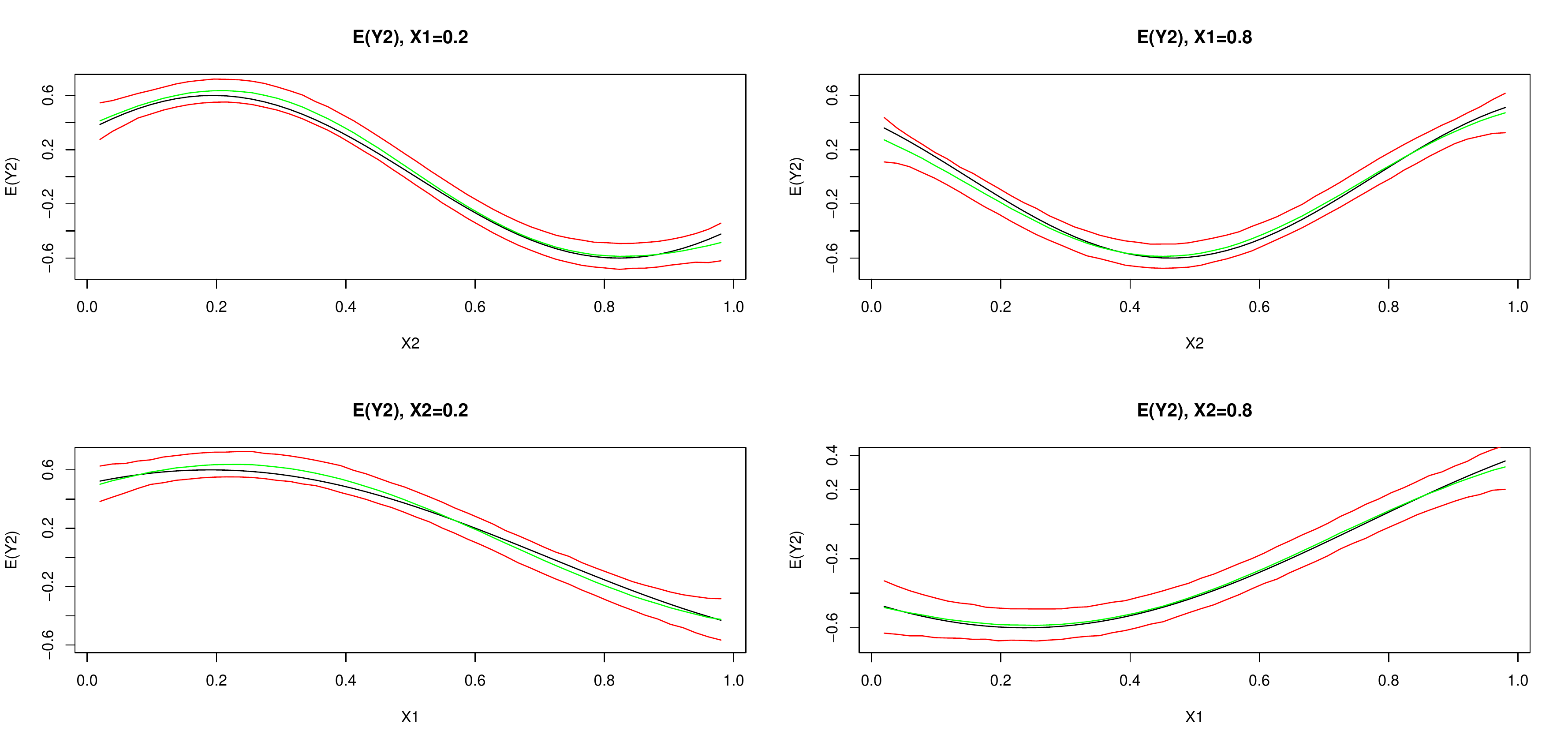}
\caption {
\textbf{Sc1}: Estimation of marginal means. The leftmost 2 columns  show the accuracy for predicting $Y_{1}$ and the rightmost 2 columns show the results for predicting $Y_{2}$.  The black and green lines represent the true and  estimated relationships, respectively. The red lines are the  limits of the pointwise 95\%  credible  intervals obtained under the true Clayton family. }
\label{fig:part1_marg}
\end{center}
\end{figure}

The simulation results show that 
\textbf{Sc1} and \textbf{Sc2} performed similarly. Since the
calibration function in {\bf Sc1} is more complicated, for the sake of reducing the paper's length we present only 
results for that scenario. The trace-plots, autocorrelation functions
and histograms of posterior samples of $\beta$, $\sigma^2_1$ and $\sigma^2_2$ are shown 
in Figure~\ref{fig:part1_trace} when the fitted copula belongs to  the correct Clayton  family (red line is the true value). 
Next we show predictions for the marginals means with 95\% credible
intervals. Since these are 2-dimensional we estimate `slices' from this
surface at values $0.2$ and $0.8$, so that
we first fix $x_1=0.2$ then $x_1=0.8$ and
similarly for $x_2$. The results are in Figure~\ref{fig:part1_marg}
(black is true, green is estimation, red are credible intervals).

\begin{figure}[!ht]
\begin{center}
\includegraphics[height=7cm,width=16cm]{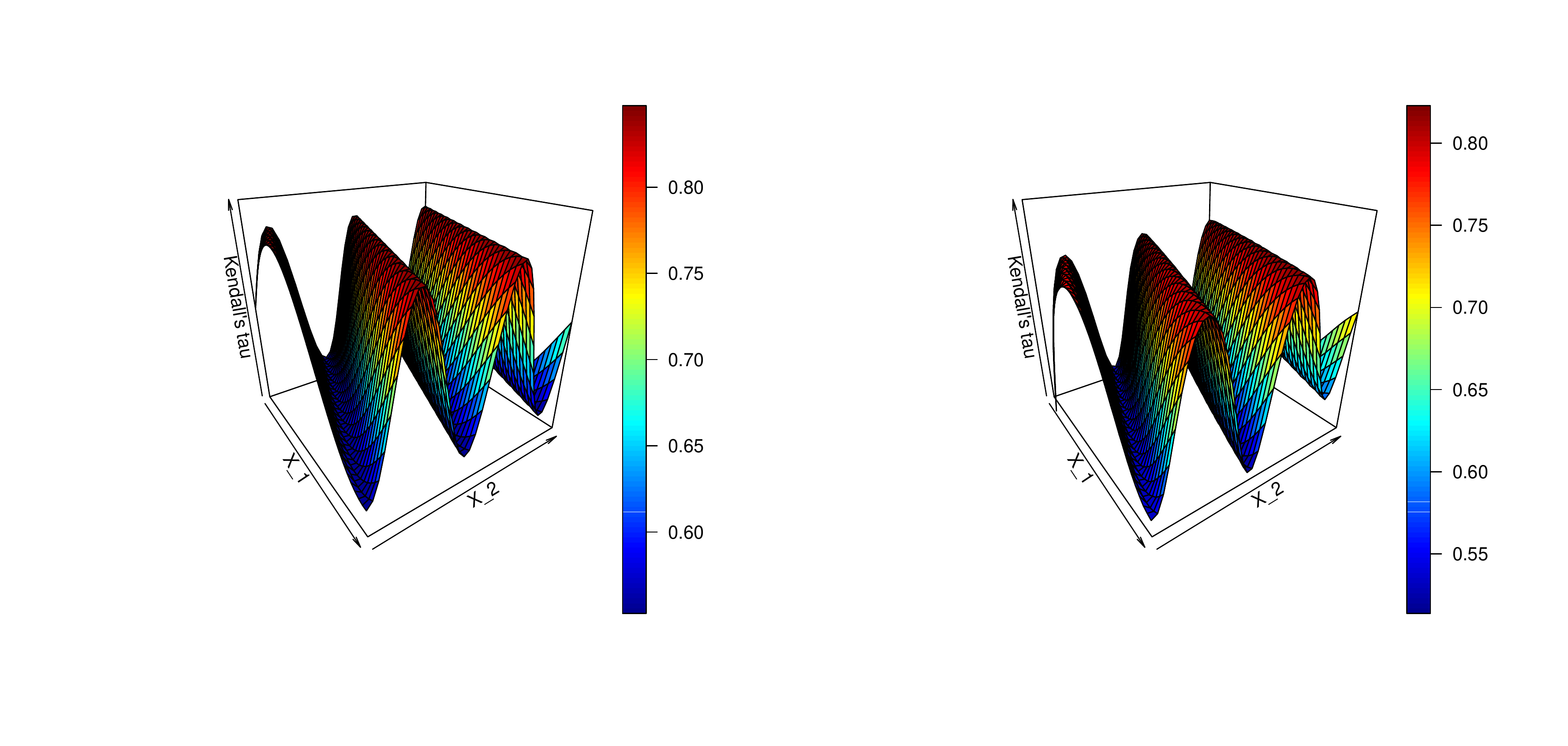}
\caption {
\textbf{Sc1}: Estimation of Kendall's $\tau$ dependence surface. The true surface (left panel) is very similar to the estimated one (right panel).   }
\label{fig:part1_surface}
\end{center}
\end{figure}

One of the inferential goals is the prediction of calibration function
or, equivalently, Kendall's $\tau$ function. In this case we are dealing with only two covariates so their joint effect can be visualized  via the calibration surface.  In Figure \ref{fig:part1_surface} we show the true calibration surface on the left panel and the fitted one on the right. The accuracy is remarkable and we are hard put to see major differences between the two panels.

\begin{figure}[!ht]
\begin{center}
\includegraphics[scale=0.45]{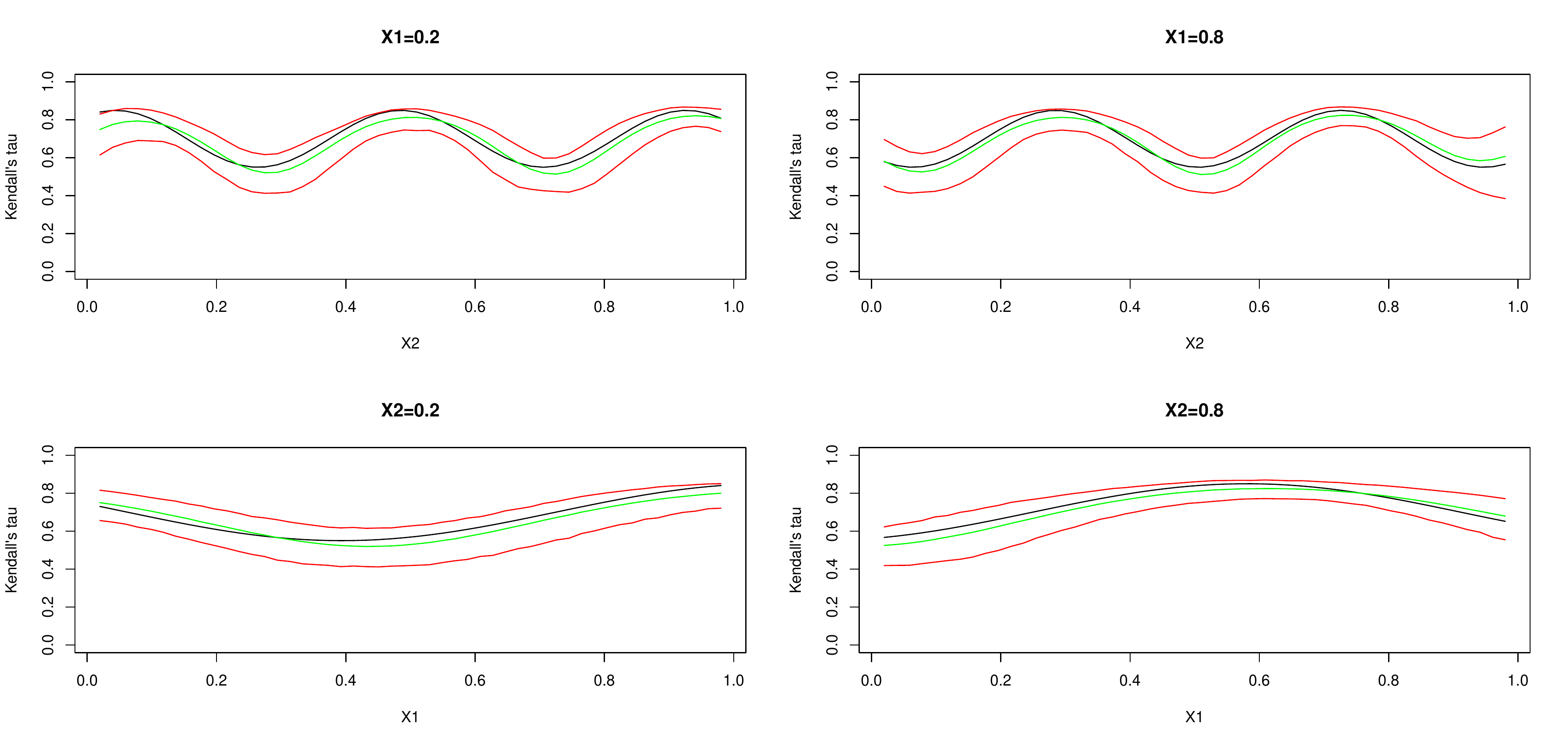}
\caption {
\textbf{Sc1}: Estimation of Kendall's $\tau$ one-dimensional projections when $X_{1}=0.2\hspace{1ex} \mbox{or}\hspace{1ex} 0.8$   (top panels) and when  $X_{2}=0.2 \hspace{1ex} \mbox{or}\hspace{1ex} 0.8$ bottom panels. The black and green lines represent the true and  estimated relationships, respectively. The red lines are the  limits of the pointwise 95\%  credible in intervals obtained under the true Clayton family.  }
\label{fig:part1_calib}
\end{center}
\end{figure}

Since  the visual comparison  of the three-dimensional true and fitted surfaces may be misleading, we also  estimate
one dimensional slices at values $0.2$ and $0.8$ and the results, shown in Figure~\ref{fig:part1_calib}, confirm the accuracy of the fit.  

\begin{figure}[!ht]
\begin{center}
\includegraphics[scale=0.2]{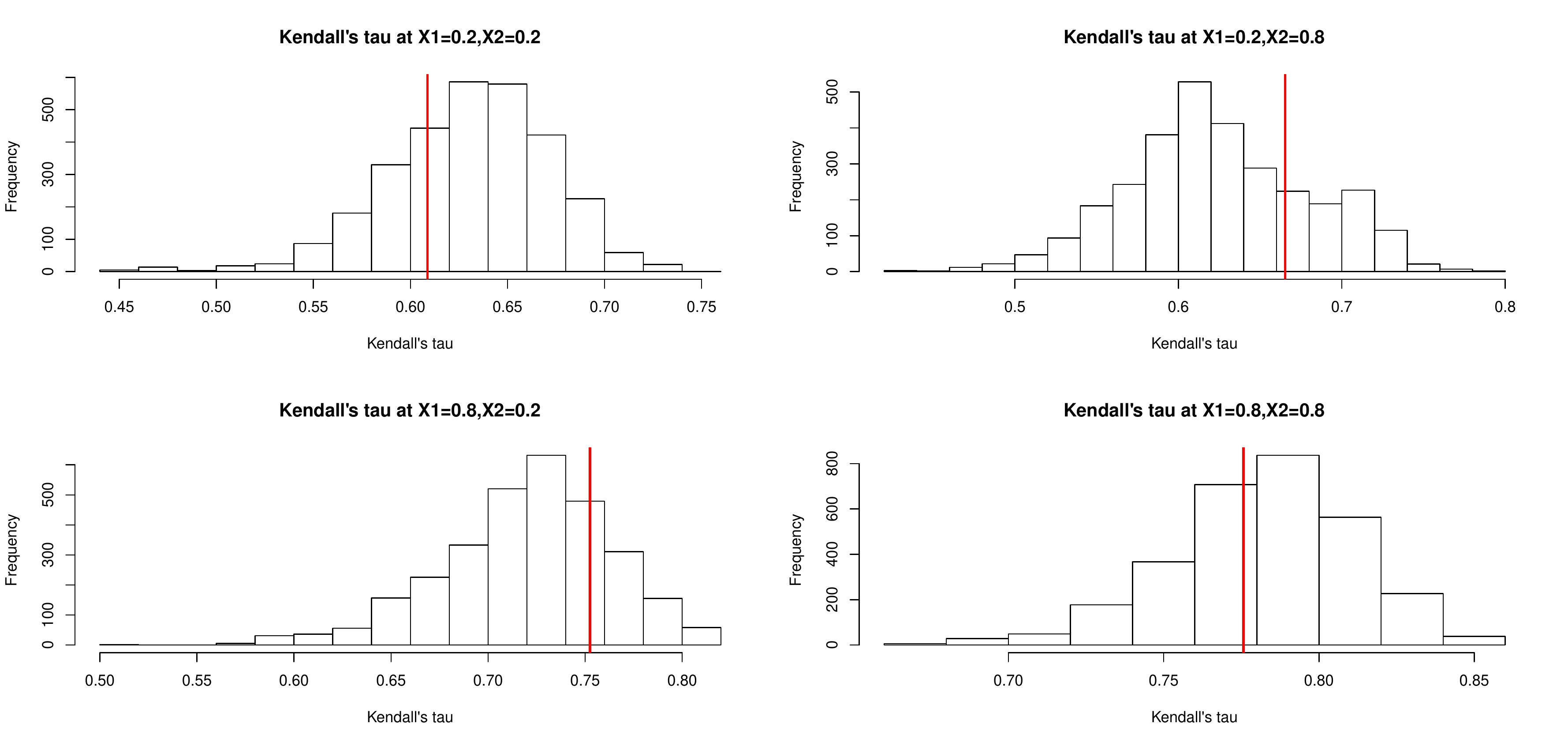}
\caption {
\textbf{Sc1}: Histogram of predicted Kendall's $\tau$ values obtained under the true Clayton copula. }
\label{fig:part1_tau}
\end{center}
\end{figure}

\begin{figure}[!ht]
\begin{center}
\includegraphics[scale=0.2]{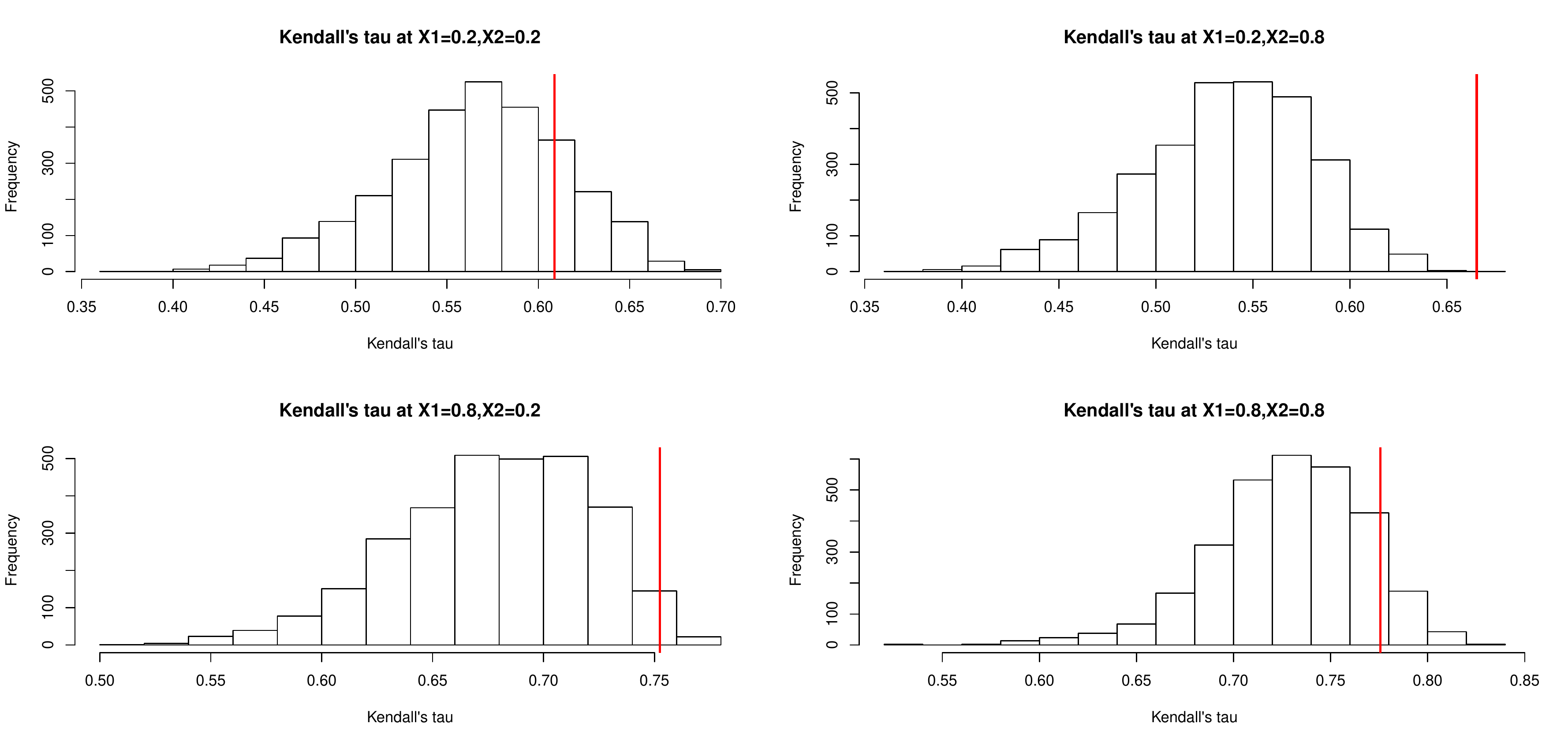}
\caption {
\textbf{Sc1}: Histogram of predicted $\tau$s (Gaussian copula) }
\label{fig:part1_tau_gauss}
\end{center}
\end{figure}

Another way to evaluate how well the model makes predictions is
to fix 4 covariate points and estimate corresponding
Kendall's $\tau$ values: $\hat \tau(0.2,0.2),\hat\tau(0.2,0.8),\hat\tau(0.8,0.2),
\hat\tau(0.8,0.8)$.  At each MCMC iteration these predictions are calculated 
and histograms (Figure~\ref{fig:part1_tau}) are constructed (red lines
are true value of $\tau$).
The same 
estimates are presented in Figure~\ref{fig:part1_tau_gauss} when the Gaussian copula is used for inference. One can notice that the 
estimates are biased in this instance, thus emphasizing the importance of identifying  the right copula family. Similar patterns have been observed when using the Frank copula. 

\begin{figure}[!ht]
\begin{center}
\includegraphics[scale=0.45]{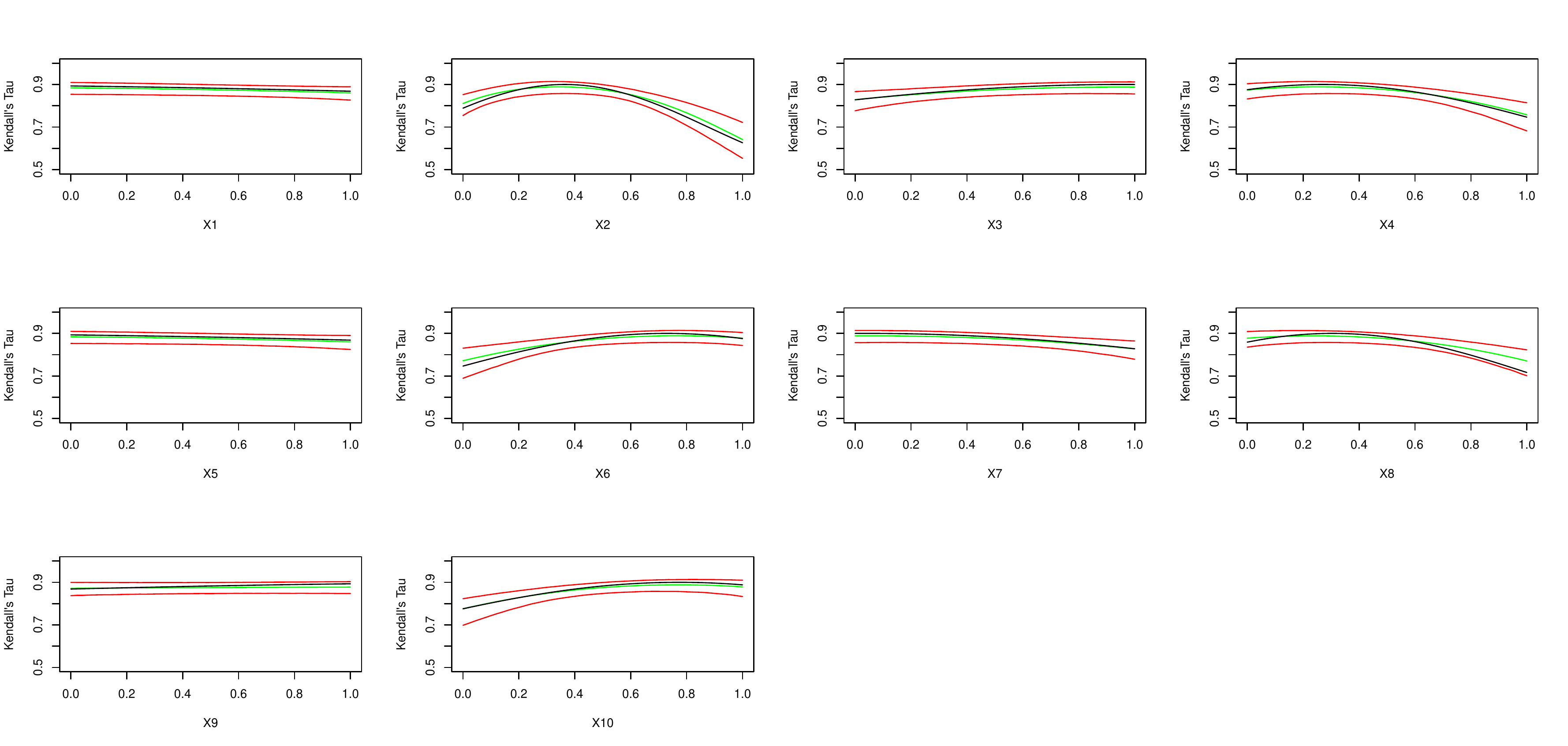}
\caption {
\textbf{Sc3}: Estimation of Kendall's $\tau$ one-dimensional projections for each coordinate fixing all other coordinates at 0.5 levels. The black and green lines represent the true and  estimated relationships, respectively. The red lines are the  limits of the pointwise 95\%  credible in intervals obtained under the true Clayton family.  }
\label{fig:part1_calib_10}
\end{center}
\end{figure}

We also show how well the algorithm estimates calibration function when covariate dimension is large. 
Figure~\ref{fig:part1_calib_10} shows one dimensional slices of Kendall's $\tau$ function for \textbf{Sc3}
which is estimated by Clayton GP-SIM model.
Each plot is produced by varying one coordinate from 0 to 1 while fixing all other coordinates at $x=0.5$. We observe that
even in this case the estimated curves are very close to true Kendall's $\tau$ function.

Finally, we focus on the accuracy of CVML, CCVML and WAIC in selecting the correct model. Table~\ref{table:part1_CVML} shows the values for each
scenario and model. Bold values indicate largest CVML/CCVML and
smallest WAIC values for each scenario.
\begin{table} [!ht]
\begin{center}
\scalebox{0.60}{
\begin{tabular}{c|| c c c |||| c|| c c c }
\hline
                               & CVML            & CCVML            & WAIC             &                                & CVML             & CCVML           & WAIC           \\ \hline
\multicolumn{1}{c}{Scenario 1} &                 &                  &                  & \multicolumn{1}{c}{Scenario 4} &                  &                 &                 \\ \hline \hline
Clayton                        & $\mathbf {496}$ & $\mathbf {417}$  & $\mathbf{-994}$  & Clayton                        & $322$            &$254$            & $-644$          \\
Frank                          & $435$           &$362$             & $-870$           & Frank                          & $277$            & $209$           & $-549$          \\
Gaussian                       & $405$           & $329$            & $-817$           & Gaussian                       & $276$            & $207$           & $-547$          \\
Clayton-Const                  & $455$           &$378$             & $-910$           & Clayton-Const                  & $\mathbf{323}$   &$\mathbf{255}$   & $\mathbf{-647}$ \\ \hline
\multicolumn{1}{c}{Scenario 2} &                 &                  &                  & \multicolumn{1}{c}{Scenario 5} &                  &                 &                 \\ \hline \hline
Clayton                        & $\mathbf{166}$  & $\mathbf{103}$   & $\mathbf{-333}$  & Clayton                        & $\mathbf{324}$   & $\mathbf{277}$  & $\mathbf{-650}$ \\  
Frank                          & $144$           & $82$             & $-289$           & Frank                          & $256$            & $216$           & $-513$           \\
Gaussian                       & $146$           & $84$             & $-293$           & Gaussian                       & $260$            & $214$           & $-520$           \\
Clayton-Const                  & $121$           & $60$             & $-243$           & Clayton-Const                  & $299$            & $257$           & $-600$            \\ \hline 
\multicolumn{1}{c}{Scenario 3} &                 &                  &                  & \multicolumn{1}{c}{Scenario 6} &                  &                 &                   \\ \hline \hline
Clayton                        & $\mathbf{613}$  &$\mathbf{536}$    & $\mathbf{-1237}$ & Clayton                        & $\mathbf{286}$   & $\mathbf{242}$  & $\mathbf{-573}$   \\   
Frank                          & $562$           & $491$            & $-1126$          & Frank                          & $216$            & $179$           & $-432$            \\
Gaussian                       & $494$           & $417$            & $-1002$          & Gaussian                       & $205$            & $165$           & $-410$             \\
Clayton-Const                  & $537$           & $462$            & $-1076$          & Clayton-Const                  & $283$            & $238$           & $-567$             \\ \hline 

\end{tabular}
}
\caption{
 CVML, CCVML and WAIC values for each Scenario and Model }
\label{table:part1_CVML}
\end{center}
\end{table}
Observe that all bold values for \textbf{Sc1},
\textbf{Sc2}, \textbf{Sc3}, \textbf{Sc5}, \textbf{Sc6}, point to the  Clayton family,
while for \textbf{Sc4}  they indicate the Clayton family with a constant calibration. We note that the correct copula is selected even when the generative calibration model is additive.


\subsubsection{Simulation Results based on multiple Replicates}

So far, the results reported were based on a single implementation of the method. In order to facilitate interpretation, we perform 50 independent replications under each of the six scenarios described previously. However, since the focus of the inference is on the copula, we shorten the simulation time by assuming that the marginals are uniform.


The MCMC sampler was run for 20000 iterations for \textbf{Sc3} and 5000 iterations for other scenarios. 
As before, the first half of iterations was ignored as a burn-in period.
For each data set, 4 estimations were done with Clayton, Frank,
Gaussian and constant Clayton copulas. For \textbf{Sc5} and \textbf{Sc6} we also
fitted the Clayton copula with an  additive model a la \cite{sabeti2014additive} for each data set. The goal is to estimate
integrated  squared Bias ($\mbox{IBias}^2$), Variance (IVar) and mean squared error (IMSE)
of Kendall's $\tau$ evaluated at covariates $X=(x_1,\ldots,x_n)^T$.
To calculate these quantities for any scenario and any
model we do the following: for each data set, point estimations
are produced $\hat {\tau}_r(x_i)$ where $r$ runs from 1 up to number
of replicates ($R$) and $i=1\ldots n$. The formulas
for $\mbox{IBias}^2$, IVar and IMSE are given by:
\begin{equation}
\begin{split}
&\mbox{IBias}^2=\sum_{i=1}^{n}\left(\frac{\sum_{r=1}^{r=R}\hat \tau_r(x_i)}{R}-\tau(x_i)\right)^2/n, \\
&\mbox{IVar}=\sum_{i=1}^{n}Var_r(\hat \tau_r(x_i))/n, \\
&\mbox{IMSE}=\mbox{IBias}^2+\mbox{IVar}.   
\end{split}
\end{equation}
We will apply these concepts not only for Kendall's $\tau$ but also
for $E(U_1|U_2=u_2,X=x)$ for different $u_2$ and $x$ combinations.
Here we use $U$s instead of $Y$s to indicate that we assume uniform
marginal distributions. \\ \\
\textbf{Results} \\
$\mbox{IBias}^2$, IVar and IMSE for each scenario and
each model are shown in Table~\ref{table:part2_tau_mse} (bold values show
smallest IMSE for each scenario).
\begin{table} [!ht]
\begin{center}
\scalebox{0.65}{
\begin{tabular}{c|| c c c| c c c|c c c| c c c }
\hline
& \multicolumn{3}{c|}{Clayton} & \multicolumn{3}{c|}{Frank} &\multicolumn{3}{c|}{Gaussian} & \multicolumn{3}{c}{Clayton Constant} \\ \hline
Scenario & $\sqrt{\mbox{IBias}^2}$ & $\sqrt{\mbox{IVar}}$ & $\sqrt{\mbox{IMSE}}$ & $\sqrt{\mbox{IBias}^2}$ & $\sqrt{\mbox{IVar}}$ & $\sqrt{\mbox{IMSE}}$ & $\sqrt{\mbox{IBias}^2}$ & $\sqrt{\mbox{IVar}}$ & $\sqrt{\mbox{IMSE}}$ & $\sqrt{\mbox{IBias}^2}$ & $\sqrt{\mbox{IVar}}$ & $\sqrt{\mbox{IMSE}}$  \\ \hline \hline
\textbf{Sc1} & 0.0223 &  0.0556 &   \textbf{0.0599}  &   0.0491  &  0.0714 &   0.0867   &  0.0664 & 0.0741 & 0.0995 & 0.1071 & 0.0133 &  0.1079  \\
\textbf{Sc2} & 0.0160  & 0.0576  &  \textbf{0.0598}   &  0.0839  &  0.0938  &  0.1258   &  0.0383 & 0.0738 & 0.0832    &  0.2208 & 0.0304 &  0.2229  \\
\textbf{Sc3} & 0.0392  & 0.0689  &  \textbf{0.0792}   &  0.0494  &  0.0912  &  0.1037   &  0.1070 & 0.1063 & 0.1509    &  0.1302 & 0.0109 &  0.1306  \\
\textbf{Sc4} & 0.0061  & 0.0318  &   0.0324   &  0.0300  &  0.0467   &  0.0555    & 0.0483  & 0.0552  &  0.0734   & 0.0028  & 0.0116 &  \textbf{0.0119}  \\
\textbf{Sc5} & 0.0712  & 0.0742  &  \textbf{0.1029}   &  0.0717  &  0.1030 &   0.1255   &  0.0793 & 0.0898 & 0.1199   &   0.1593 & 0.0217  & 0.1607  \\
\textbf{Sc6} & 0.0286  & 0.0602  &  \textbf{0.0667}  &   0.0406  &  0.0834  &  0.0928  &  0.0540 & 0.0720 & 0.0901   &   0.0935 & 0.0160 &  0.0949 \\
\hline
\end{tabular}
}
\caption{
Estimated $\sqrt{\mbox{Bias}^2}$, $\sqrt{\mbox{IVar}}$ and $\sqrt{\mbox{IMSE}}$ of Kendall's $\tau$ for each Scenario and Model}
\label{table:part2_tau_mse}
\end{center}
\end{table}
Note that the smallest IMSE is produced when fitting the correct  model and  copula family. 
For each simulated data set and each model, $E(U_1|U_2=u_2,X=x)$ were
estimated. For all scenarios except for \textbf{Sc3} we let each $x_1,x_2,u_2$
to take values in the set $\{0.2,0.4,0.6,0.8\}$,
making a total of 64 combinations. For \textbf{Sc3} we let $u_2$ to take
values from $\{0.2,0.4,0.6,0.8\}$, while $x$ can take 33 values scattered in $[0,1]^{10}$,
making a total of 132 combinations.  
The results are presented in Table~\ref{table:part2_EY_mse}.
\begin{table} [!ht]
\begin{center}
\scalebox{0.65}{
\begin{tabular}{c|| c c c| c c c|c c c| c c c }
\hline
& \multicolumn{3}{c|}{Clayton} & \multicolumn{3}{c|}{Frank} &\multicolumn{3}{c|}{Gaussian} & \multicolumn{3}{c}{Clayton Constant} \\ \hline
Scenario & $\sqrt{\mbox{IBias}^2}$ & $\sqrt{\mbox{IVar}}$ & $\sqrt{\mbox{IMSE}}$ & $\sqrt{\mbox{IBias}^2}$ & $\sqrt{\mbox{IVar}}$ & $\sqrt{\mbox{IMSE}}$ & $\sqrt{\mbox{IBias}^2}$ & $\sqrt{\mbox{IVar}}$ & $\sqrt{\mbox{IMSE}}$ & $\sqrt{\mbox{IBias}^2}$ & $\sqrt{\mbox{IVar}}$ & $\sqrt{\mbox{IMSE}}$  \\ \hline \hline
\textbf{Sc1} &  0.0038 & 0.0131   &  \textbf{0.0137}   &  0.0279    &   0.0132  &  0.0309    &   0.0338 & 0.0163 & 0.0375   &  0.0237 & 0.0034 & 0.0240  \\
\textbf{Sc2} & 0.0034  & 0.0145   &  \textbf{0.0149}   &  0.0256    &   0.0310 &   0.0403    &   0.0226 & 0.0184 & 0.0292  &   0.0644 & 0.0084 & 0.0649 \\
\textbf{Sc3} & 0.0068  & 0.0148   &  \textbf{0.0163}   &  0.0117    &   0.0175 &   0.0211    &   0.0320 & 0.0262 & 0.0413  &   0.0198 & 0.0024 & 0.0199 \\
\textbf{Sc4} & 0.0015  & 0.0070   & 0.0072   & 0.0265    &0.0076 &0.0275    &0.0346  & 0.0122 & 0.0367  & 0.0007  &0.0028 & \textbf{0.0029} \\
\textbf{Sc5} & 0.0170 &  0.0192  &  \textbf{0.0257}  &   0.0347 &    0.0250 &  0.0428    &   0.0398 & 0.0209 & 0.0450   &  0.0479 & 0.0069 & 0.0484  \\
\textbf{Sc6} & 0.0076  & 0.0158   &  \textbf{0.0176}   &  0.0372    &   0.0189 &   0.0417    &   0.0384 & 0.0171 & 0.0420   &  0.0219 & 0.0050 & 0.0225 \\
\hline
\end{tabular}
}
\caption{
Estimated $\mbox{IBias}^2$, IVar and IMSE of $E(U_1|U_2,X)$ for each Scenario and Model}
\label{table:part2_EY_mse}
\end{center}
\end{table}

Focusing on \textbf{Sc5} and \textbf{Sc6}, the estimated integrated bias, variance and MSE
for Kendall's $\tau$ and $E(U_2|U_2=u_2,X=x)$ when fitting Clayton GP-SIM and true
Clayton Additive model are shown in Table~\ref{table:part2_tau_mse_add}.
We observe that even though Clayton GP-SIM has best IMSE among other copula families
it performs worse than Clayton additive model which generated data sets.
\begin{table} [!ht]
\begin{center}
\scalebox{0.80}{
\begin{tabular}{c|| c c c| c c c }
\hline
\multicolumn{1}{c}{}& \multicolumn{6}{c}{Kendall's Tau} \\ \hline
& \multicolumn{3}{c|}{Clayton GP-SIM} & \multicolumn{3}{c}{Clayton Additive}  \\ \hline
Scenario & $\sqrt{\mbox{IBias}^2}$ & $\sqrt{\mbox{IVar}}$ & $\sqrt{\mbox{IMSE}}$ & $\sqrt{\mbox{IBias}^2}$ & $\sqrt{\mbox{IVar}}$ & $\sqrt{\mbox{IMSE}}$  \\ \hline \hline
\textbf{Sc5} & 0.0712  & 0.0742  &  0.1029   &  0.0528  &  0.0469 &   \textbf{0.0707} \\
\textbf{Sc6} & 0.0286  & 0.0602  &  0.0667  &   0.0069  &  0.0390  &  \textbf{0.0396} \\ \hline
\multicolumn{1}{c}{}& \multicolumn{6}{c}{$E(U_2|U_1,X)$} \\ \hline
& \multicolumn{3}{c|}{Clayton GP-SIM} & \multicolumn{3}{c}{Clayton Additive}  \\ \hline
Scenario & $\sqrt{\mbox{IBias}^2}$ & $\sqrt{\mbox{IVar}}$ & $\sqrt{\mbox{IMSE}}$ & $\sqrt{\mbox{IBias}^2}$ & $\sqrt{\mbox{IVar}}$ & $\sqrt{\mbox{IMSE}}$  \\ \hline \hline
\textbf{Sc5} & 0.0170 &  0.0192  &  0.0257  &   0.0136 &    0.0105 &  \textbf{0.0172}  \\
\textbf{Sc6} & 0.0076  & 0.0158   &  0.0176   &  0.0018    &   0.0095 &   \textbf{0.0097} \\
\hline

\end{tabular}
}
\caption{
Estimated $\sqrt{\mbox{Bias}^2}$, $\sqrt{\mbox{IVar}}$ and $\sqrt{\mbox{IMSE}}$ of Kendall's $\tau$ and $E(U_1|U_2,X)$ for
GP-SIM and Additive models}
\label{table:part2_tau_mse_add}
\end{center}
\end{table}

Finally we show how well CVML and WAIC perform in choosing
correct model. For selecting between different copula families or to check whether dependence is covariate-free we
just pick the model with largest CVML or smallest WAIC. 
Table~\ref{table:part2_CVML} shows how often Clayton model is selected over
other models using CVML and WAIC for \textbf{Sc1}, \textbf{Sc2}, \textbf{Sc3}, \textbf{Sc5} and \textbf{Sc6}.
Similarly, Table~\ref{table:part2_CVML_const} shows how often Clayton-constant is selected over
other models for \textbf{Sc4}.
\begin{table} [!ht]
\begin{center}
\scalebox{0.6}{
\begin{tabular}{c|| c c  |c c | c c }
\hline
& \multicolumn{2}{c|}{Frank} &\multicolumn{2}{c|}{Gaussian}   & \multicolumn{2}{c}{Clayton Constant} \\ \hline
Scenario                     & CVML  & WAIC   & CVML  & WAIC  & CVML  &  WAIC \\ \hline \hline
\textbf{Sc1}                 & 100\% & 100\%  & 100\% & 100\% & 98\%  & 98\%   \\
\textbf{Sc2}                 & 100\% & 100\%  & 100\% & 100\% & 100\% & 100\%   \\
\textbf{Sc3}                 & 100\% & 100\%  & 100\% & 100\% & 100\% & 100\%   \\
\textbf{Sc5}                 & 100\% & 100\%  & 100\% & 100\% & 100\% & 100\%   \\
\textbf{Sc6}                 & 100\% & 100\%  & 100\% & 100\% & 90\%  & 90\%   \\
\hline
\end{tabular}
}
\caption{ The percentage of correct decisions for each selection criterion when comparing the correct Clayton model with a non-constant calibration with all the other models:  Frank model with non-constant calibration, Gaussian model with non-constant calibration, Clayton model with non-constant calibration. Notice that the CCVML and the CVML criteria are exactly equal in the case in which the marginals are uniform.  
}
\label{table:part2_CVML}
\end{center}
\end{table}

\begin{table} [h]
\begin{center}
\scalebox{0.6}{
\begin{tabular}{c|| c c |c c | c c  }
\hline
                               & \multicolumn{2}{c|}{Clayton} & \multicolumn{2}{c|}{Frank} & \multicolumn{2}{c}{Gaussian} \\ \hline
Scenario                       & CVML & WAIC                  & CVML  & WAIC               & CVML  & WAIC                  \\ \hline \hline
\textbf{Sc4}                   & 78\% & 78\%                  & 100\% & 100\%              & 100\% & 100\%   \\
\hline
\end{tabular}
}
\caption{ The percentage of correct decisions for each selection criterion when comparing the correct Clayton model with a constant calibration with all the other models: Clayton model with non-constant calibration, Frank model with non-constant calibration and the Gaussian model with non-constant calibration. Notice that the CCVML and the CVML criteria are exactly equal in the case in which the marginals are uniform.
}
\label{table:part2_CVML_const}
\end{center}
\end{table}
We can conclude that both selection measures perform similarly for all scenarios. Also, selection results
show that choosing between copula families is easy, while CVML and WAIC do not perform exceptionally well
in selecting between different forms of calibration
function (GP-SIM and SA). 
Since \textbf{Sc5} and \textbf{Sc6} where simulated with Clayton additive calibration, we show
how often Clayton Additive model is selected over Clayton GP-SIM using different criteria (Table~\ref{table:part2_CVML_add}).
\begin{table} [!ht]
\begin{center}
\scalebox{0.80}{
\begin{tabular}{c|| c c  }
\hline
              & \multicolumn{2}{c}{Clayton GP-SIM}  \\ \hline
Scenario      & CVML & WAIC \\ \hline \hline
\textbf{Sc5}  & 94\% & 94\%    \\
\textbf{Sc6}  & 70\% & 70\%     \\
\hline
\end{tabular}
}
\caption{ The percentage of correct decisions for each selection criterion when comparing the correct additive model with GP-SIM with non-constant calibration
}
\label{table:part2_CVML_add}
\end{center}
\end{table}
Again CVML and WAIC perform similarly. The poor performance of criteria for
\textbf{Sc6} is not that surprising since the additive calibration in this scenario
has almost SIM form as functions $y=x$ and $y=x^2$ are similar for 
$x\in [0,1]$. 


\subsection{Red Wine Data}

We consider the  data of \cite{cortez2009modeling} consisting of various
physicochemical tests of 1599 red variants of the Portuguese "Vinho Verde" wine.
Acidity and density are  properties closely associated with the quality of wine and grape, respectively. Of interest here is to study the dependence pattern between`fixed acidity' ($Y_{fa}$) and `density' ($Y_{de}$) and how it changes with values
of other variables: `volatile acidity', `citric acid', `residual sugar',
`chlorides', `free sulfur dioxide', `total sulfur dioxide', `pH', `sulphates'
and `alcohol', denoted $X_{va},X_{ca},X_{rs},X_{ch},X_{fs},X_{ts},X_{ph},X_{su},X_{al}$,
respectively. Response variable are linearly transformed
to have mean 0 and standard deviation of 1, similarly covariates where transformed
to be between 0 and 1. \\
To select the appropriate copula family, we fit GP-SIM with `Clayton', `Frank', `Gaussian',
`Gumbel' and `T-3' (student T with 3 degrees of freedom) dependencies.
For each model the MCMC was run for 10000 iterations with 5000 burn-in period. We used 30 inducing inputs for the marginals and  calibration function estimation
($m_1=m_2=m=30$). The resulting CVML, CCVML and WAIC values are shown in Table~\ref{table:wine}.
\begin{table} [!ht]
\begin{center}
\scalebox{0.7}{
\begin{tabular}{r|| r r r r r   }
\hline
& Clayton & Frank & Gaussian & Gumbel & T-3  \\ \hline
CVML            &	   -1858	&	 -1816	&	\textbf{-1788}	&	 -1829	&	 -1810	\\
CCVML	        &	    -582	&	  -547	&	\textbf{-522}	&	  -558	&	  -534	\\
WAIC      	&	    3713	&	  3634	&	\textbf{3572}	&	  3656	&	  3621	\\
\hline
\end{tabular}
}
\caption{ Red Wine data: CVML, CCVML and WAIC criteria values different models   
}
\label{table:wine}
\end{center}
\end{table}

All model selection measures indicate that  among candidate copula families the most suitable one is the Gaussian one.  The GP-SIM coefficients ($\beta$) fitted under the Gaussian copula family are shown in Table \ref{table:wine_beta}. 

\begin{table}[!ht]
\begin{center}
\scalebox{0.8}{
\begin{tabular}{r | r  r }
\hline
Variable      & Posterior Mean  &  95\% Credible Interval \\ \hline
$X_{va}$      & $0.274$         & $[0.154,  0.389]$ \\
$X_{ca}$      & $-0.336$        & $[-0.413,   -0.254]$ \\
$X_{rs}$      & $-0.076$        & $[-0.278,   0.271]$ \\ 
$X_{ch}$      & $0.060$         & $[-0.246,   0.259]$ \\ 
$X_{fs}$      & $0.276$         & $[0.106,   0.410]$ \\ 
$X_{ts}$      & $0.402$         & $[0.248,   0.608]$ \\ 
$X_{ph}$      & $0.155$         & $[0.054,   0.286]$ \\ 
$X_{su}$      & $0.501$         & $[0.342,   0.601]$ \\ 
$X_{al}$      & $0.463$         & $[0.382,   0.517]$ \\ 
\hline
\end{tabular}
}
\caption{Wine data: Posterior means and quantiles of $\beta$ }
\label{table:wine_beta}
\end{center}
\end{table}


The credible intervals suggest that not all covariates may be needed
to model dependence between responses. For example,  `residual sugars' and
`chlorides' seem to  not affect the  calibration function  so we consider a model in which they are omitted from the conditional copula model. In all models, we include  
all the covariates in the marginal distributions. 
For comparison, we have also fitted all Gaussian GP-SIM models with
only one covariate, and with no covariates at all (constant). The computational
algorithm to fit GP-SIM when the conditional copula depends on only one variable is very
similar to the one described above. The main difference is that there is no $\beta$ variable
and the inducing inputs (for calibration function) are evenly spread on $[0,1]$. The testing
results are shown in Table~\ref{table:wine_selection}. 
\begin{table} [!ht]
\begin{center}
\scalebox{0.7}{
\begin{tabular}{c|| c c c  }
\hline
Variables                                               & CVML          & CCVML                 &  WAIC  \\ \hline \hline
ALL	                                                &	-1788	&	 -522		&	 3572	\\   \hline
$X_{va},X_{ca},X_{fs},X_{ts},X_{ph},X_{su},X_{al}$	&	-1805	&	 -532		&	 3608	\\
$X_{va}$	                                        &	-1823	&	 -552		&	 3646	\\
$X_{ca}$	                                        &	-1815	&	 -541		&	 3629	\\
$X_{rs}$	                                        &	-1849	&	 -582		&	 3698	\\
$X_{ch}$	                                        &	-1842	&	 -578		&	 3688	\\
$X_{fs}$	                                        &	-1852	&	 -584		&	 3705	\\
$X_{ts}$	                                        &	-1851	&	 -583		&	 3700	\\
$X_{ph}$	                                        &	-1816	&	 -557		&	 3633	\\
$X_{su}$	                                        &	-1841	&	 -571		&	 3682	\\
$X_{al}$	                                        &	-1847	&	 -577		&	 3697	\\
Constant	                                        &	-1849	&	 -584	        &	3700	\\
\hline
\end{tabular}
}
\caption{ Wine data: CVML, CCVML and WAIC criteria values for variable selection in conditional copula 
}
\label{table:wine_selection}
\end{center}
\end{table}

\begin{figure}[!ht]
\begin{center}
\includegraphics[scale=0.45]{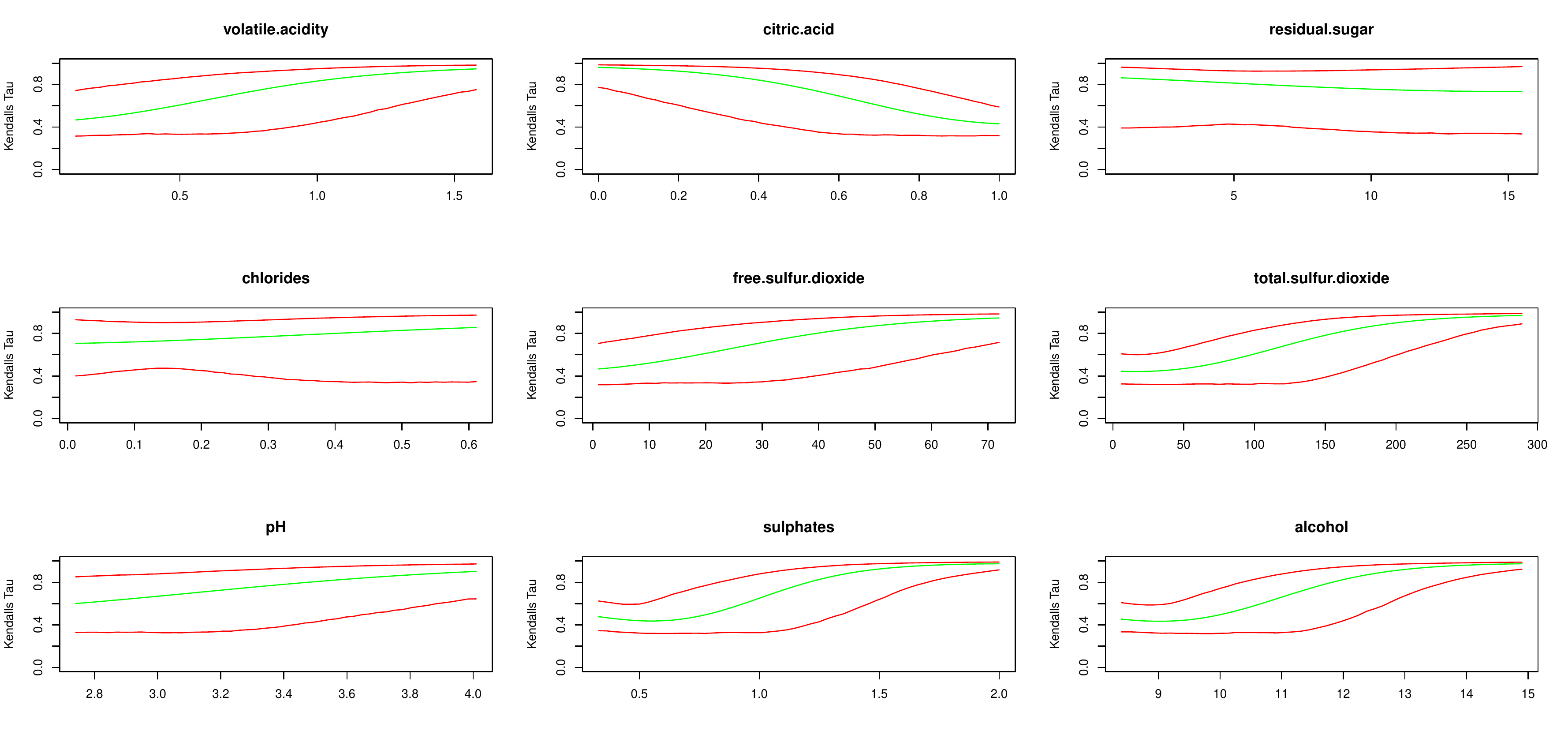}
\caption {Wine Data: Slices of predicted Kendall's $\tau$ as function of covariates. Red curves represent 
95\% credible intervals }
\label{fig:wine_tau}
\end{center}
\end{figure}

Based on the selection criteria results we  conclude that all nine covariates are required to
explain the dependence structure of two responses.
Figure~\ref{fig:wine_tau} shows 1-dimensional plots of  Kendall's $\tau$ calibration
curve with 95\% credible as a function of covariates. The plots are constructed by
varying one predictor while fixing all others at their mid-range values.  

The plots clearly demonstrate that when covariates are fixed at their mid-range values,
the conditional correlation between `fixed acidity' and `density' increases with `volatile acidity',
`free sulfur dioxide', `total sulfur dioxide', `pH', `sulphates' and
`alcohol', and  decreases with levels of `citric acid'. These relationships can influence the preparation method of the wine.

\begin{figure}[!ht]
\begin{center}
\includegraphics[scale=0.45]{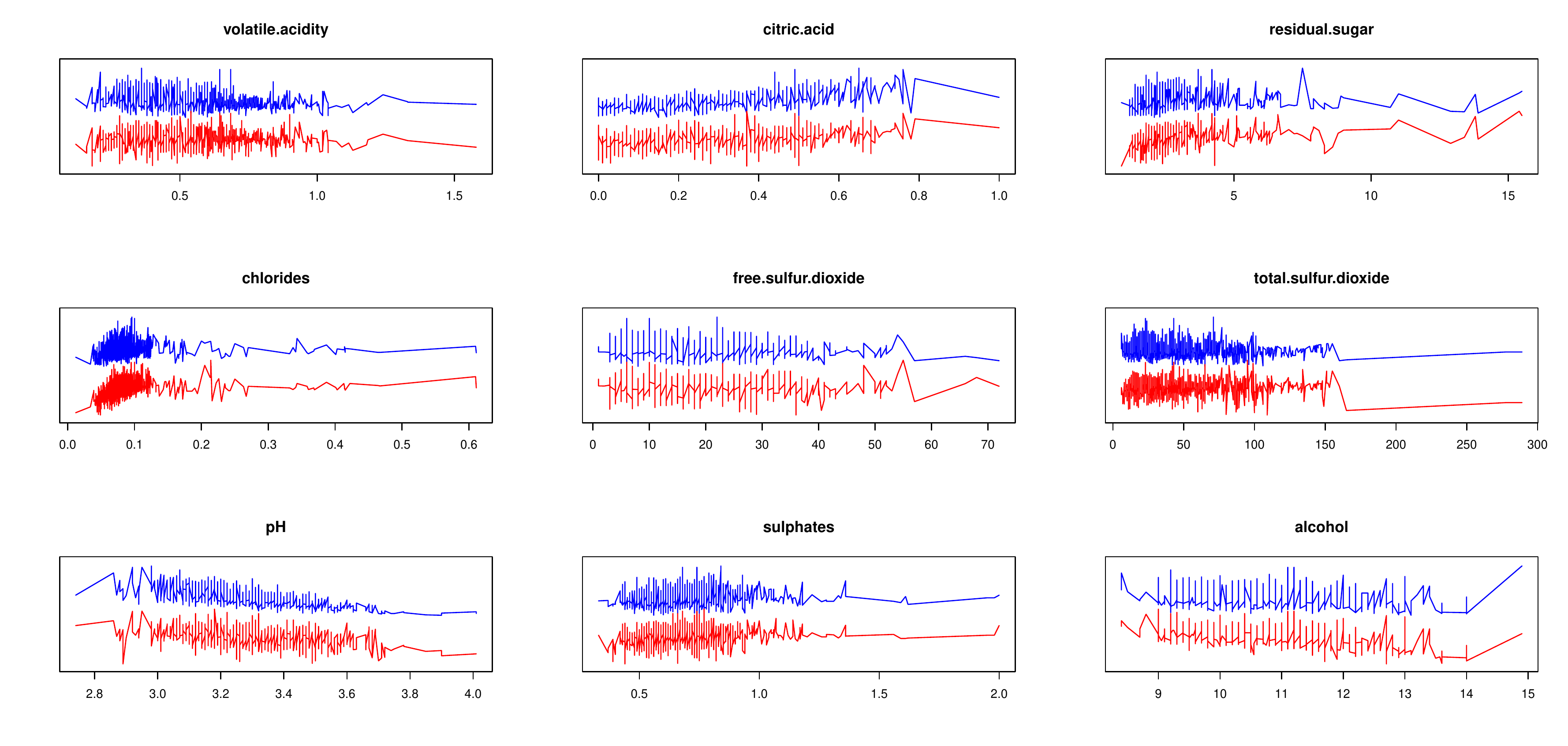}
\caption {Wine Data: Plots of `fixed acidity'(blue) and `density'(red) (linearly transformed to fit on one plot) against covariates.
 }
\label{fig:wine_trace}
\end{center}
\end{figure}


In order to demonstrate the difficulty one would have in gauging the complex evolution of dependence between two responses as a function of covariates we plot in Figure~\ref{fig:wine_trace} the response variables together as they vary with each covariate. It is clear that the model manages to identify a pattern that would be very difficult to distinguish without the help of a flexible mathematical model.\\

\section{Simplifying Assumption}

\subsection{Model Misspecification and the Simplifying Assumption}
\label{cop-mis}
Understanding whether  the data support the  SA or not  is usually important for the subject matter analysis since a  dependence structure that does not depend on the covariates can be of scientific interest. The SA has also a serious impact on  the statistical analysis, because it has the potential to  simplify greatly the estimation of the copula.  There is however, an interesting connection between model misspecification and SA.

\begin{figure}[!ht]
\begin{center}
\includegraphics[scale=0.45]{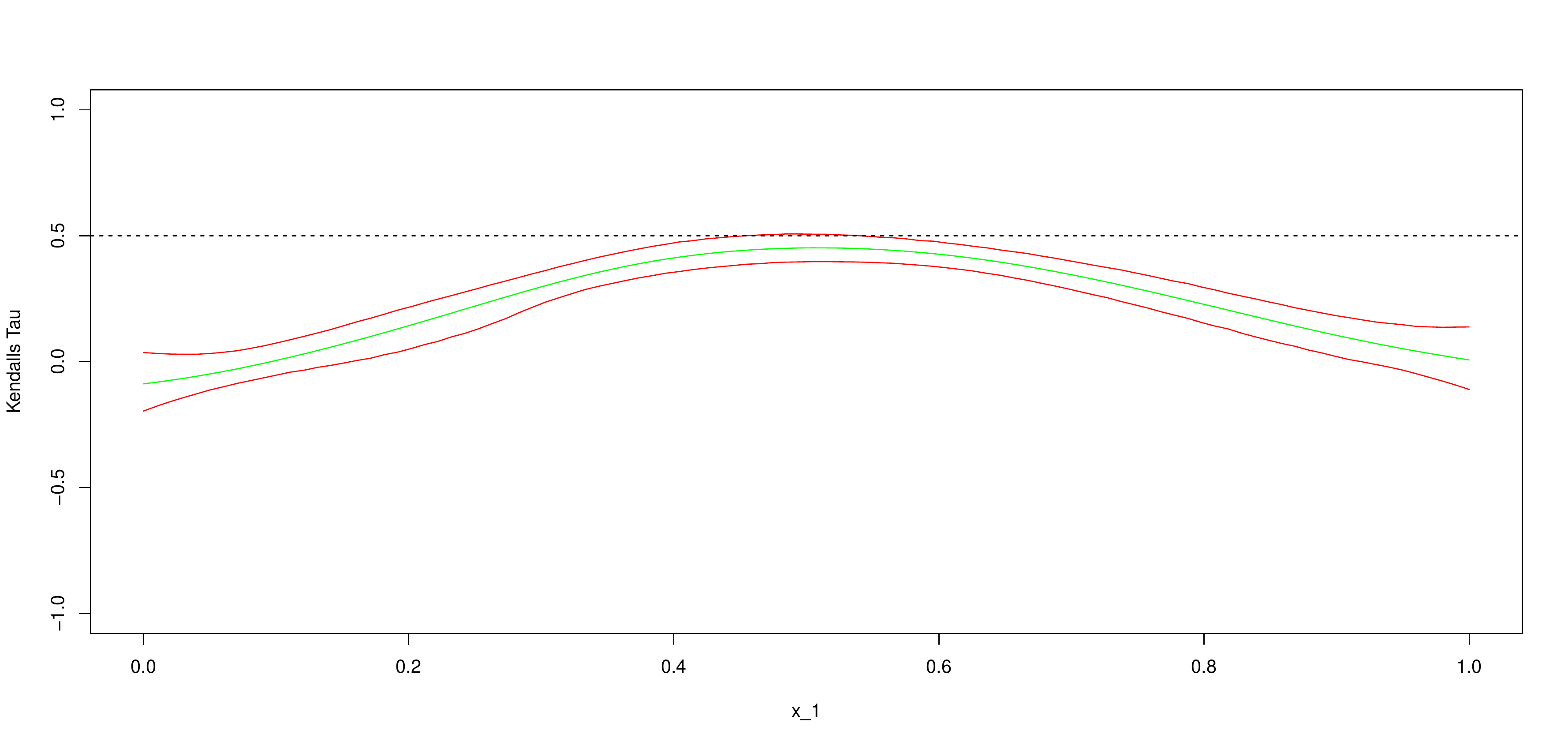}
\caption {
Estimation of Kendall's $\tau$ as a function of $x_1$ when only first covariate is used in estimation. The black and green lines represent the true and  estimated relationships, respectively. The red lines are the  limits of the pointwise 95\%  credible in intervals obtained under the true Clayton family.  }
\label{fig:tau_missed_cov}
\end{center}
\end{figure}

To illustrate the point, we consider two independent random variables, $X_1,X_2$ to serve as covariates in the  Clayton copula model in which  SA is satisfied, the sample size $n=1500$ and
\begin{itemize}
\item[]    $f_1(x)=0.6\sin(5x_1+x_2),$\\
           $f_2(x)=0.6\sin(x_1+5x_2),$\\
           $\tau(x)=0.5,$\\
           $\sigma_1=\sigma_2=0.2.$ 
\end{itemize}
When we fit a GP-SIM model with the correct Clayton copula family, but with the $X_2$  covariate  omitted from both marginal and copula models,   the  estimated Kendall's $\tau(X_1)$  exhibits a clear non-constant shape, as
seen in Figure~\ref{fig:tau_missed_cov}. 
The CVML, CCVML and WAIC criteria, whose values are shown in Table~\ref{table:missed_cov_sel}, unanimously vote for a nonconstant calibration function.

\begin{table} [!ht]
\begin{center}
\scalebox{0.7}{
\begin{tabular}{c|| c c c  }
\hline
Variables             & CVML          & CCVML         & WAIC  \\ \hline \hline
$X_1$                 &	-508	&	-174	&		1017	\\   \hline
Constant	      &	-570	&	-232	&		1140	\\
\hline
\end{tabular}
}
\caption{ Missed covariate: CVML, CCVML and WAIC criteria values for model with conditional copula depends on one covariate and when it is constant. 
}
\label{table:missed_cov_sel}
\end{center}
\end{table}

While one may expect a nonconstant pattern  when the two covariates are dependent, this  residual effect of $X_1$  on the copula may be surprising when $X_1$ and $X_2$ are independent. 

We can gain some understanding by considering a simplified example in which $Y_i|X_1,X_2 \sim N(f_i(X_1,X_2),1)$ for $i=1,2$, and $\rcov (Y_1,Y_2|X_1,X_2)= \rcorr(Y_1,Y_2|X_1,X_2)=\rho$, hence independent of $X_1$ and $X_2$.
When considering marginal models that include only $X_1$, thus leading to  residuals $W_i=Y_i-E[Y_i|X_1]$ for $i=1,2$, we are interested in understanding why 
$\rcov (W_1,W_2|X_1)$  is not constant in $X_1$. 
Using the statistical properties of covariance along with the properties of  conditional expectation one can show
\beq
\rcov (W_1,W_2|X_1)=\rcov (Y_1,Y_2|X_1),
\label{sa-cov}
\eeq
and
\begin{align}
 \rcov (Y_1,Y_2|X_1)&=E[\rcov (Y_1,Y_2|X_1,X_2)]+\rcov (E[Y_1|X_1,X_2],E[Y_2|X_1,X_2])\nonumber\\
 &=\rho + \rcov (f_1(X_1,X_2),f_2(X_1,X_2)), 
 \label{sa-missp}
 \end{align}
 where the covariance in \eqref{sa-missp} is with respect to the distribution of $X_2$. Hence it is apparent that the conditional covariance $\rcov (W_1,W_2|X_1)$ will generally not be constant in $X_1$.  It should be noted that if the true means have additive form, i.e. $f_i(X_1,X_2)=\bar f_i(X_1)+ \tilde  f_i(X_2)$, for $i=1,2$, then the covariances in \eqref{sa-cov} are indeed constant in $X_1$, but the estimated value of $\rcov(Y_1,Y_2|X_1)$ will be biased. Although here we focused on the covariance as a measure of dependence, the argument is extendable to copula parameters or Kendall's tau, but the calculations are more involved.

In conclusion, violation of the SA may be  due to the omission of important  covariates from the model. This phenomenon along with the knowledge that in general  it is difficult to measure all the variables with potential effect on the dependence pattern, suggests that a non-constant copula is a prudent choice.


\subsection{A Permutation-based Criterion to Detect Data Support for the  Simplified Assumption}

In this section we propose a to modify the CVML and the conditional CCVML  method to identify data support for 
SA after the  copula family is selected. 

As was shown in previous sections, the selection criteria
included in the paper do not perform well when true calibration is constant. This is in line with 
\cite{cra-sabeti} who also noted that the traditional Bayesian model selection criteria, e.g. the Deviance information criterion (DIC) of \cite{spie:best:etal:02}, tend to prefer  the more complex calibration model  over 
 a simple model with  constant calibration even when the latter is actually correct. 
 In addition of the simulations presented in the previous section, we add here that when the  marginal distributions are estimated, the performance of the existing criteria worsens. To illustrate,  we have simulated 50 replicates
of sample sizes 1500 using Clayton copula from \textbf{Sc1}, \textbf{Sc4} and \textbf{Sc5}. 
Each sample is fitted with the general model introduced here and a constant Clayton copula, while 
marginals are  estimated using a general GP.
Table~\ref{table:SA_1} shows the
proportion of correct decisions for the
three scenarios and various selection criteria. These results show that even for 
a large sample size, the proportion of right decisions  for \textbf{Sc4}, i.e. when SA holds, is
quite low. One of the explanations is that the general model does a good job at capturing the constant trend of the calibration function and yields predictions that are not too far from the ones produced with the simpler (and correct) model.  The modified CVML  we propose is inspired by two 
desiderata: i) to separate the set of observations used for prediction from the set of observations used for fitting the model, and ii) to amplify the impact of  the copula-induced errors in the CCVML calculation.  The former will reduce the implicit bias one gets when the same data is used for estimation and testing, while the latter is expected to increase the power to identify SA. 

For i) we randomly  partition  the data into a  training set $\mathcal D=\{Y_{1i},Y_{2i},X_i\}_{i=1,\ldots,n}$
and a test set $\mathcal D^*=\{Y_{1i}^*,Y_{2i}^*,X_i^*\}_{i=1,\ldots,n^*}$. In our numerical experiments we have kept  two thirds  of
observations in  the training set. In order to achieve ii) we note that permuting the response indexes will not affect the  copula term if SA is indeed satisfied and will perturb the prediction when SA is not satisfied. However, one must cautiously implement this idea, since the permutation $\lambda: \{1,\ldots,n^*\} \rightarrow \{1,\ldots,n^*\}$ will affect the marginal model fit, regardless of the SA status,  as $Y_{j\lambda(i)}$  will be paired with $X_i$, for all $j=1,2$.  Below we describe the  permutation-based CVML criterion that combines i) and ii).

Assume that the fitted GP-SIM model yields posterior
samples from the conditional distribution of  latent variables and parameters $w^{(t)} \sim w|\mathcal D$, $t=1\ldots M$. Then we define the  observed data criterion   
as the predictive log probability of the test cases which can be easily estimated from posterior samples,
as follows:
\begin{equation}
\begin{split}
&\mbox{CVML}_{obs}= \sum_{i=1}^{n^*} \log P(Y_{1i}^*,Y_{2i}^*|\mathcal D,X_i^*)\approx \sum_{i=1}^{n^*} \log \left\{ \frac{1}{M} \sum_{t=1}^{M} P(Y_{1i}^*,Y_{2i}^*|w^{(t)},X_i^*)  \right\} = \\
&           = \sum_{i=1}^{n^*} \log \left\{ \frac{1}{M} \sum_{t=1}^{M} \frac{1}{\sigma_1^{(t)}}\phi\left( \frac{Y_{1i}^*-f_{1i}^{*(t)}}{\sigma_1^{(t)}} \right) \frac{1}{\sigma_2^{(t)}}\phi\left( \frac{Y_{2i}^*-f_{2i}^{*(t)}}{\sigma_2^{(t)}} \right) \times \right . \nonumber\\
& \left . \times c\left[ \Phi\left( \frac{Y_{1i}^*-f_{1i}^{*(t)}}{\sigma_1^{(t)}} \right), \Phi\left( \frac{Y_{2i}^*-f_{2i}^{*(t)}}{\sigma_2^{(t)}} \right) | \theta_i^{*(t)} \right ]   \right\},                                     
\end{split}
\end{equation}
where $f_{1i}^{*(t)},f_{2i}^{*(t)},\theta_i^{*(t)}$ are the predicted values for the test cases produced by the GP-SIM model. 
 
 Consider $J$ permutations of $\{1\ldots n^*\}$ which we denote as $\lambda_1,\ldots,\lambda_J$, and compute
$J$ permuted CVMLs as:
\begin{eqnarray}
\mbox{CVML}_j &=& \sum_{i=1}^{n^*} \log \left \{ \frac{1}{M} \sum_{t=1}^{t=M} \frac{1}{\sigma_1^{(t)}}\phi\left( \frac{Y_{1i}^*-f_{1i}^{*(t)}}{\sigma_1^{(t)}} \right) \frac{1}{\sigma_2^{(t)}}\phi\left( \frac{Y_{2i}^*-f_{2i}^{*(t)}}{\sigma_2^{(t)}} \right) \right. \times \nonumber\\
& \times & \left . c\left[ \Phi\left( \frac{Y_{1i}^*-f_{1i}^{*(t)}}{\sigma_1^{(t)}} \right), \Phi\left( \frac{Y_{2i}^*-f_{2i}^{*(t)}}{\sigma_2^{(t)}} \right) | \theta_{\lambda_j(i)}^{*(t)} \right ]   \right \}. 
\label{cvml-perm}                                 
\end{eqnarray}
Note that $CVML_{obs}$ differs from $CVML_j$  only in the values of the copula parameters. While for the former we use $\theta(X^*_i)$, in the latter we use $\theta(X^*_{\lambda_j(i)})$ for the dependence between $Y^*_{1i}$ and $Y^*_{2i}$.   If calibration is constant
then $\mbox{CVML}_{obs}$ and $\mbox{CVML}_{j}$ should be similar, hence we define the evidence
\begin{equation} 
\mbox{EV}=2 \times \min \left \{ \frac{\sum \limits_{j=1}^{J} \mathbbm{1}_{\{ CVML_{obs}<CVML_j \}}}{J},\frac{\sum\limits_{j=1}^{J} \mathbbm{1}_{\{ CVML_{obs}>CVML_j \}}}{J} \right\}.
\label{pval}
\end{equation}

Under the null model with constant calibration with known marginals  and if we assume that $CVML_{obs}$ and $\{CVML_j: 1\le j \le J\}$ are iid for each $j$, then each term inside the $\min$ function in \eqref{pval} has a $\ru(0,1)$ limiting distribution when $J \rightarrow \infty$.  In that case it follows that 
$P(EV < 0.05) =0.05$.  In practice, the ideal situation just described is merely an approximation since the $\{CVML_j: 1\le j \le J\}$ are not independent and we compute EV using a fixed number of permutations. Nevertheless, the ideal setup can be used  to build our decision that  when  $EV>0.05$  the data support SA, and  otherwise they do not. 

A similar rule can be build using the   CCVML criterion. For instance, its value for test data is  
\begin{equation}
\mbox{CCVML}_{obs}= \frac{1}{2}\sum_{i=1}^{n^*} \log P(Y_{1i}^*|\mathcal D,X_i^*,Y_{2i}^*) + \frac{1}{2}\sum_{i=1}^{n^*} \log P(Y_{2i}^*|\mathcal D,X_i^*,Y_{1i}^*).
\label{ccvml-obs}
\end{equation}
The permutation-based version of \eqref{ccvml-obs} can be obtained using the same principle as in \eqref{cvml-perm} thus leading to the counterpart of \eqref{pval} for CCVML.

Table~\ref{table:SA_2} shows the proportion of correct decisions using proposed methods with 1000 and 500 samples in training and test set
respectively, and $J=500$ permutations. The results, especially those  for  \textbf{Sc4},  clearly show an important improvement in the rate of making the 
correct selection, with only a small decrease in the power to detect non-constant calibrations. We can also notice that CVML and CCVML performed similarly.

\begin{table} [h]
\centering
\begin{minipage}{.48\textwidth}
\begin{center}
\scalebox{1}{
\begin{tabular}{c|| c c c  }
Scenario      & CVML  &  CCVML  & WAIC        \\ \hline
\textbf{Sc1}  & 100\% &  100\%  & 100\%        \\
\textbf{Sc4}  & 74\%  &  78\%   & 74\%        \\
\textbf{Sc5}  & 100\% &  100\%  & 100\%       \\ \hline
\end{tabular}
}
\caption{ The percentage of correct decisions for each selection criterion and scenarios.
GP-SIM and SA were fitted with Clayton copula, sample size is 1500 }
\label{table:SA_1}
\end{center}
\end{minipage}\hfill
\begin{minipage}{.48\textwidth}
\begin{center}
\scalebox{1}{
\begin{tabular}{c|| c c   }
Scenario      & CVML  &  CCVML          \\ \hline
\textbf{Sc1}  & 98\%  &  96\%          \\
\textbf{Sc4}  & 92\%  &  90\%           \\
\textbf{Sc5}  & 100\% &  100\%         \\ \hline
\end{tabular}
}
\caption{ The percentage of correct decisions for each selection criterion and scenario.
Predicted CVML and CCVML values based on $n=1000$ training  and $n^*=500$  test data, respectively. The calculation of EV is based on a random sample of $500$ permutations.
 }
\label{table:SA_2}
\end{center}
\end{minipage}\hfill
\end{table}

\section{Conclusion and Future Work}

The inclusion of a dynamic copula in the model comes with a significant computational price.  The inclusion can be justified by the need for an exploration of dependence, or because it can improve the predictive accuracy of the model.   The simplifying assumption is often used as a way to bypass the need for a conditional copula model. However, we have showed that  even if the simplifying assumption holds for the true model,   when we ignore the contribution of one covariate, fitted copula is no longer constant. 

We have proposed a Bayesian procedure  to estimate the 
calibration function of a conditional copula model jointly with  the marginal distributions. In our attempt to move away from an additive model hypothesis we consider a sparse Gaussian process priors used in conjunction with a single index model. The resulting procedure  reduces the dimensionality  of the parameter space and can be used for small and moderate covariate
dimension.  

We have introduced a couple  of selection criteria to help select the copula family from a set of candidates and to gauge data support in favour of the simplifying assumption. While the former task seems to be achieved by all criteria considered, the latter is a particularly difficult problem  and we are excited about the good performance exhibited by our permutation-based version of the cross-validated marginal likelihood criterion.  Its theoretical properties are the focus of our ongoing work and we plan to extend its use  to identifying those covariates that do not influence the calibration function.

\section*{Acknowledgement}

We thank Keith Knight and Stanislav Volgushev for helpful suggestions that have improved the paper.
Funding  support of this work was provided by  the Natural Sciences and Engineering Research Council of Canada and the Canadian Statistical Sciences Institute.

%

\end{document}